\documentclass[useAMS,usenatbib]{mn2e}
\usepackage{aas_macros}
\usepackage{amssymb}
\usepackage{setspace}
\usepackage[pdftex]{graphicx}
\usepackage{url}

\newcommand {\apgt} {\ {\raise-.5ex\hbox{$\buildrel>\over\sim$}}\ }
\newcommand{\appropto}{\mathrel{\vcenter{
  \offinterlineskip\halign{\hfil$##$\cr
    \propto\cr\noalign{\kern2pt}\sim\cr\noalign{\kern-2pt}}}}}

\title[Cloud angular momentum and effective viscosity]{Cloud angular momentum and effective viscosity in global SPH simulations with feedback}
\author[David J. Williamson, Robert J. Thacker, James Wurster, and Brad K. Gibson]{David J. Williamson$^{1,2,3}$\thanks{E-mail:
david-john.williamson.1@ulaval.ca}, Robert J. Thacker$^{3}$, James Wurster$^{3,4}$, and Brad K. Gibson$^{3,5}$\\
$^{1}$D\'{e}partement de physique, de g\'{e}nie physique et d'optique, Universit\'{e} Laval, Qu\'{e}bec, QC, G1V 0A6, Canada\\
$^{2}$Centre de Recherche en Astrophysique du Qu\'{e}bec, Qu\'{e}bec, QC, Canada\\
$^{3}$Department of Astronomy and Physics, Saint Mary's University, Halifax B3H 3C3, Canada\\
$^{4}$Monash Centre for Astrophysics (MoCA) and School of Mathematical Sciences, Monash University, Clayton, VIC 3800, Australia\\
$^{5}$Jeremiah Horrocks Institute, University of Central Lancashire, Preston PR1 2HE, UK}
\begin{document}

\date{Accepted, June 2014}

\pagerange{\pageref{firstpage}--\pageref{lastpage}} \pubyear{2099}

\maketitle

\label{firstpage}

\begin{abstract}
We examine simulations of isolated galaxies to analyse the effects of localised feedback on the formation and evolution of molecular clouds. Feedback contributes to turbulence and the destruction of clouds, leading to a population of clouds that is younger, less massive, and with more retrograde rotation. We investigate the evolution of clouds as they interact with each other and the diffuse ISM, and determine that the role of cloud interactions differs strongly with the presence of feedback: in models without feedback, scattering events dramatically increase the retrograde fraction, but in models with feedback, mergers between clouds may slightly increase the prograde fraction.  We also produce an estimate of the viscous time-scale due to cloud-cloud collisions, which increases with increasing strength of feedback ($t_\nu\sim20$ Gyr vs $t_\nu\sim10$ Gyr), but is still much smaller than previous estimates ($t_\nu\sim1000$ Gyr); although collisions become more frequent with feedback, less energy is lost in each collision than in the models without feedback.
\end{abstract}


\section{Introduction}

Giant molecular clouds (GMCs) are a fundamental component of galactic structure, making an important contribution to the ISM, and having a dominant role in hosting star formation. Mergers between GMCs may also act as an effective viscosity that is weak but not negligible \citep[hereafter WT12]{2012MNRAS.421.2170W}. The mergers that generate this viscosity also contribute to the evolution of the internal structure of the simulated GMCs, including the orientation of the GMCs' spins. The precise impact of feedback on the GMC population is also an unresolved question. It is thus important to perform galaxy-scale simulations to properly understand the impact of feedback and cloud-cloud interactions on the GMC population, as well as the impact of the GMC population on galactic evolution.

Increasing computing power has permitted galaxy-scale hydrodynamic simulations with sufficient resolution and simulation time to resolve molecular cloud evolution \citep{2008MNRAS.391..844D,2008ApJ...680.1083R,2008MNRAS.385.1893D,2009MNRAS.392..294A,2009ApJ...700..358T,2011MNRAS.417.1318D,2011MNRAS.413.2935D,2011ApJ...730...11T,2012MNRAS.425.2157D,2013MNRAS.432..653D,2013MNRAS.tmpL.173D,2013ApJ...776...23B}. These models typically consist of a smooth exponential disc of gas which fragments into clouds as the system evolves. In our previous paper (WT12) we performed numerical simulations of this type and were able to determine the strength of effective viscosity resulting from cloud-cloud collisions in these models. We found the viscous time-scale is on the order of $t_\nu\sim10$ Gyr, much shorter than previous estimates of $t_\nu\sim1000$ Gyr \citep{2002ApJ...581.1013B}. The viscous time-scale in these simulations is on the order of a Hubble time, but while this suggests that the effective viscosity due to cloud-cloud collisions is not a dominant effect, it is still considerably stronger than previously predicted. The viscous timescale may be even shorter at low resolutions, perhaps having a significant effect on cosmological simulations that insufficiently resolve the disc.

However, it has been noted that the properties and evolution of molecular clouds depend strongly on the choice of numerical models and parameters, such as the model for the stellar potential \citep{2012MNRAS.425.2157D}, the strength, nature, and presence of stellar feedback \citep{2011ApJ...730...11T,2011MNRAS.413.2935D,2012MNRAS.425.2157D}, softening length, and temperature floor (WT12). Our previous numerical simulations (WT12) were performed in the absence of feedback, but feedback is known to have a significant effect on the properties and evolution of clouds. This will likely have an effect on the effective viscosity from cloud collisions, as the details of cloud interactions will depend on cloud substructure, and on the velocity distribution, which will be directly affected by energy input from feedback.

Stellar feedback is traditionally performed by adding thermal or kinetic energy into regions that pass certain criteria for star formation \citep[as discussed in numerous 
places e.g.][]{2000ApJ...545..728T,2006MNRAS.373.1074S,2009ApJ...695..292C,2010ApJ...717..121C}, and simultaneously transferring some mass from the gaseous component of the simulation into the collisionless stellar component. Sub-grid models have also been produced that model feedback with effective equations of state \citep[e.g. ][]{2003MNRAS.339..289S}. These methods typically assume that star formation is not well resolved in time or space, and so it is possible (and indeed necessary) to make use of simplified models that represent the large-scale average effects of feedback - e.g. the input of thermal or kinetic energy - without accounting for the particular details of the star formation and feedback processes. For instance, it is not possible to directly capture the photoheating of H\textsc{II} regions by O/B associations if the spatial resolution is insufficient to resolve H\textsc{II} regions and the numerical time-step is larger than the typical lifetimes of O/B stars, and hence such small-scale effects must be included as sub-grid models, if included at all. In this work, we use the feedback method of \citet{2000ApJ...545..728T}, which assumes that feedback is dominated by supernovae, that large stars produce supernovae immediately, and that each star particle contains the entire stellar initial mass function. We note that these assumptions are not necessary when applied to simulations with resolutions sufficient to resolve individual molecular clouds, and that more sophisticated methods \citep[e.g.][]{2012MNRAS.422.2609R,2013MNRAS.433...69H} could improve the accuracy of our simulated galaxy and its molecular clouds.

To further quantify the effects of feedback in simulated discs, we examine the source of the angular momentum distribution of clouds. Observations \citep{1999A&AS..134..241P,2003ApJ...599..258R,2011ApJ...732...79I,2011ApJ...732...78I} have shown that $40-60$\% of molecular clouds spin retrograde with respect to Galactic rotation. An unperturbed disc with a falling rotation curve should primarily form prograde clouds, unless the clouds form in a contrived geometry \citep{1966MNRAS.131..307M,1993prpl.conf..125B}. Large-scale perturbations such as spiral shocks \citep{1995MNRAS.275..209C} may potentially drive the production of retrograde clouds \citep[or at least affect the cloud population;][]{2014MNRAS.439..936F}, although previous simulations \citep{2011MNRAS.417.1318D,2009ApJ...700..358T,2011ApJ...730...11T,2013ApJ...776...23B} that have produced retrograde clouds have successfully done so without a galactic spiral. The fraction of retrograde clouds varies greatly between these simulations, and so the source of the angular momentum distribution has remained unclear.

\citet{2009ApJ...700..358T} performed simulations where the first clouds that formed from the galaxy's initially smooth density profile were strongly prograde, with retrograde clouds forming at later time ($t>140$ Myr) from over-dense gas already disturbed by cloud interactions -- \citet{2013ApJ...776...23B} found $18\%$ of clouds were retrograde after $240$ Myr. \citet{2008MNRAS.391..844D} similarly states that retrograde clouds form as a result of clouds forming from an inhomogeneous ISM, stirred by cloud collisions and/or feedback, but finds that $\sim40$\% of clouds are retrograde. There is a further disagreement in whether retrograde fractions increase \citep{2011MNRAS.417.1318D} or decrease \citep{2011ApJ...730...11T} with increasing strength of feedback.

In this work we compare the properties of molecular clouds as a function of angular momentum to resolve these discrepancies and determine the prime drivers of the angular momentum distribution and the effects of feedback. The structure of the paper is as follows: in section~\ref{simsect} we present our simulations, including codes used and initial conditions. In section~\ref{analsect} we summarise our analysis techniques for identifying and tracking clouds, and for quantifying the differences between the cloud populations. In section~\ref{resultsect} we give the results of these analysis techniques, and comment on their significance. In section~\ref{concsect} we present our conclusions.

\section{Simulation}\label{simsect}

We conduct new simulations of isolated galaxies that use identical initial conditions to our lowest softening length Milky Way simulation model in WT12 (named LowSoftMW), using the OpenMP N-body AP$^3$M \citep{ap3m} SPH \citep{1992ARA&A..30..543M} 
code \textsc{hydra} \citep{phydra}. Our two new simulations differ from LowSoftMW in that they include stellar feedback. We compare these two simulations with LowSoftMW, which we refer to as the "no-feedback" run. These models consist of a stellar disc, stellar bulge and dark matter halo generated by GalactICs \citep{1995MNRAS.277.1341K, 2005ApJ...631..838W, 2008ApJ...679.1239W}, with a gas disc with the same initial scale-height and scale-length as the stellar disc, and an initial temperature of $10^4$ K. These are moderate resolution models, with $4\times10^5$ gas particles, $5\times10^5$ stellar particles, $5\times10^5$ dark matter particles, and a softening length of $60$ pc. The disc scale length is $2.81$ kpc, truncated at $30$ kpc by the complementary error-function with a scale-length of $0.1$ kpc. The scale height is initially $0.36$ kpc, and the total disc mass is $5.2\times10^{10}M_\odot$, with the gaseous disc making up $10$\% of this. This results in a gas mass resolution of $13000$ M$_\odot$ per particle. The halo mass is $7.3\times10^{11} M_\odot$. These parameters are chosen to mimic the Milky Way.

These simulations use an adaptive time-step, but in practice this quickly reaches the same approximately constant value in all simulations. This time-step is $\Delta T\approx43$ kyr. Dumps of data are produced every $20$ time-steps, or approximately once every $900$ kyr. This is the time-resolution for our tracking and analysis of clouds.

\subsection{Star formation and feedback}\label{feedbackhydra}
As noted, we use the feedback method implemented in \textsc{hydra} by \citet{2000ApJ...545..728T} and also described in \citet{2013MNRAS.431.2513W}. In this model, star formation occurs when
\begin{itemize}
\item gas is sufficiently cool and dense ($n_\mathrm{H}>10^3$ cm$^{-3}$, $T<3\times10^4$ K)
\item the flow is convergent ($\nabla\cdot\mathbf{v}<0$)
\item and the gas is partially self-gravitating ($\rho_\mathrm{g}>0.4\rho_\mathrm{DM}$).
\end{itemize} In practice, the threshold density of $10^3$ cm$^{-3}$ sets a ``soft'' density ceiling for the gas, as feedback prevents gas from collapsing to greater densities. In our feedback runs the majority of the gas mass has densities of $n_\mathrm{H}=10^{-1}$ cm$^{-3}$ to $n_\mathrm{H}=10^{3}$ cm$^{-3}$, with a low-density tail that extends to $n_\mathrm{H}=10^{-4}$ cm$^{-3}$.

A gas particle tracks its cumulative mass of stars formed calculated from a Lagrangian form of the Schmidt Law \citep{1998ApJ...498..541K}. This is tracked as a ``sub-grid'' mass of star formation until the gas and stars are explicitly decoupled by the production of a star particle (i.e. the sum of the internal stellar mass and internal gas mass of a gas particle is equal to the particle's kinematic mass). The star formation proscription is explicitly given as
\begin{equation}
\frac{\mathrm{d}M_\mathrm{*,SG}}{\mathrm{d}t} = \epsilon_\mathrm{SFR} M_\mathrm{g,SG}\sqrt{4 \mathrm{\pi} G \rho_g},
\end{equation} where ${\mathrm{d}M_\mathrm{*,SG}}/{\mathrm{d}t}$ is the rate at which the particle's sub-grid gas mass is converted into sub-grid star mass, $M_\mathrm{g,SG}$ is the remaining sub-grid gas mass, $\rho_g$ is the SPH gas density of the gas particle, and $\epsilon_{SFR}$ is the star-formation efficiency. We select $\epsilon_{SFR}=0.02$, consistent with the observed and simulated low efficiency of star formation \citep{2007ApJ...654..304K}.

Each gas particle produces up to two star particles. The first particle is produced when the cumulative sub-grid mass of star formation within a gas particle reaches half of the gas particle's initial mass. Following the production of the first particle, a second particle is produced when the particle's stellar mass has reached $80$\% of the mass of the remaining gas particle. This circumvents an issue where the cumulative star formation only asymptotically approaches the full mass of the remaining particle because the star formation rate is proportional to the remaining gas mass. This approximation implies that in this second phase of a gas particle's star formation, the last $20$\% of star formation is instantaneous. While this method reduces the computational load, it also forces the gas and stars to be dynamically coupled until the particle is produced, which has a distinct impact on dynamical evolution as it prevents stellar mass from leaving the star-forming region during its early evolution.

In the same time-step that a gas particle produced a star particle, feedback is also applied directly to all of the progenitor particle's neighbours following the SPH kernel. The feedback is applied as $\epsilon^*\times10^{51}$ erg of thermal energy per $100 M_\odot$ of stars formed, where $\epsilon^*=0.4$ is a dimensionless parameter \citep[as in][]{1993MNRAS.265..271N,2013MNRAS.431.2513W}. To prevent overcooling, each gas particle that has recently been subject to feedback uses a reduced ``effective density'' in the radiative cooling subroutine, motivated by the assumption of pressure equilibrium between the different phases of the interstellar medium (ISM). This effective density acts as a simple model for the unresolved small-scale evolution within a gas particle, and exponentially decays to the gas particle's true density. Explicitly, the radiative cooling equation becomes
\begin{equation}
e_i\rightarrow e_i - n_\mathrm{eff}^2 \Lambda(T),
\end{equation} where $e_i$ is the particle's specific energy, $\Lambda(T)$ is the cooling function at this temperature and metallicity, and $n_\mathrm{eff}$ is the effective density. The time-scale of this decay, $t_*$, is a free parameter, and effectively sets the strength of feedback, with a larger $t_*$ producing stronger feedback because the feedback energy persists in the ISM for a longer period.

\subsection{Simulations}

We present three simulations. These simulations have identical initial conditions, but differ in feedback parameters. One simulation was performed without feedback (LowSoftMW from WT12), one simulation with ``weak'' feedback ($t_*=0.5$ Myr) and one simulation with ``strong'' feedback ($t_*=1$ Myr). Both choices of $t_*$ are lower than typically selected \citep[e.g.][]{2006MNRAS.373.1074S}, and it is likely that feedback in the models presented here is weaker than in a realistic environment. At the moderate resolution of $4\times10^5$ particles, stronger feedback results in a disc that is too hot to fragment into molecular clouds, and hence it is necessary to choose parameters that result in feedback that is perhaps weaker than is realistic. As we note below, this results in clouds that are more massive and more strongly prograde than would result from more realistic parameters.

The no-feedback run has been previously presented as ``LowSoftMW'' in WT12, and does not include star formation. Without any form of feedback the star formation rate would likely become extremely large, rapidly consuming large amounts of gas and producing a large mass of star particles. In this work we are chiefly interested in the hydrodynamic effects of feedback on cloud structure, and producing a large mass of star particles would dramatically alter the dynamics of the system, making it more difficult to disentangle the effects of feedback.

Additionally, we only apply a dynamic temperature floor in the simulation without feedback. Without feedback, nothing impedes the cooling of dense gas, and the temperature of gas in molecular clouds can only be controlled artificially. As most of the mass of gas is in cold clouds, the temperature of the majority of the gas will be controlled by the cooling floor, and so the results in the no-feedback simulation will depend on the choice of temperature floor. Here we use a dynamic temperature floor of \citet{2008ApJ...680.1083R} that ensures that the Jeans mass is well-resolved. In the two feedback runs, gas is heated in a self-consistent manner by stellar feedback, and so it is not necessary to artificially impose a dynamic temperature floor.

With these three models we can contrast the effects of including and varying the strength of feedback. The differences between these models are summarised in Table~\ref{ictable}.

\begin{table}
\begin{tabular}{ccc}
\hline\hline
Name & $t_*$ (Myr) & Dynamic Temperature\\
~&~&Floor\\
\hline
No Feedback & N/A & Yes\\
Weak Feedback & 0.5 & No\\
Strong Feedback & 1.0 & No\\
\hline
\end{tabular}
\caption{\label{ictable}Summary of simulation parameters. The simulations only differ in the strength and presence of feedback, and the presence of a dynamic temperature floor.} 
\end{table}

\section{Analysis}\label{analsect}

We identified and tracked the clouds and determined the viscous time-scale using the method described in WT12. The procedure involves a modified friends-of-friends algorithm \citep{1985ApJ...292..371D} to identify clouds, and tracking particle IDs to follow clouds across time-steps. This allows us to identify clouds that are merging or separating. A ``merger'' is identified when a cloud at one data dump contains more than half the gas particles of each of two or more clouds from the previous data dump. A ``separation'' is identified when at least half of the particles from each of two or more clouds at one data dump were contained within a single cloud in the previous data dump. By identifying the change in the cloud's orbital kinetic energy across each merger or separation event, we can determine the rate of energy loss, and hence the viscous time-scale due to cloud-cloud collisions. As cloud mergers are often chaotic, several merger and separation events may be identified before the clouds completely merge, and so it is necessary to also track separation events so that cloud mergers are not over-counted. Indirect interactions from the gravity of nearby clouds or large-scale tidal torques are not captured by this method.

To quantify the differences in the evolution of the cloud systems between runs, we define two quantities, $f_{\mathrm{r},i} (t',t)$ and $n_{\mathrm{eff},i} (t',t)$. These quantities track the evolution of particles from a particular cloud (cloud $i$) at a chosen time ($t'$). These two quantities can be measured for $t<t'$ to examine the history of the gas that formed the cloud, as well as for $t>t'$ to examine the future of the gas that currently comprises the cloud.

$f_{\mathrm{r},i} (t',t)$ is the fraction of a cloud $i$'s particles (defined at time $t'$) that are also in any cloud at time $t$, i.e. $1-f_{\mathrm{r},i}$ is the fraction of cloud $i$'s particles that are now in the diffuse ISM. The subscript ``r'' stands for the fraction of cloud particles ``remaining'' at that time.

$n_{\mathrm{eff},i} (t',t)$ is the ``effective'' number of clouds that the particles from cloud $i$ at $t'$ are distributed amongst at time $t$, giving an effective number of progenitor or child clouds for this cloud. If we take the particles from cloud $i$ at $t'$ and simply count the number of clouds at time $t$ that these particles are in, we find a number that is large and rapidly varying in time. This is because a small number of particles is weighted the same amount as a large number of particles. If, for example, a single particle from a large cloud breaks off and joins another cloud, directly counting the number of clouds the parent cloud's particles are distributed between would read as if the cloud had split in two, which is not an entirely accurate description of this undramatic event.

Instead, we define $n_{\mathrm{eff},i} (t',t)$ as $n_{\mathrm{eff},i} (t',t)=1/(\Sigma_j (N_{ij}/N_{i})^2)$ across all clouds $j$, where $N_{ij}$ is the number of particles from cloud $i$ at $t=t'$ that are in cloud $j$ at time $t$, and $N_i$ is the total number of particles from cloud $i$ at $t=t'$ that are in any cloud at time $t$. This weights clouds with a greater fraction of the particles less than clouds with a smaller fraction of the particles. To demonstrates some examples, if cloud $i$ is divided into $n$ equal portions, then $n_{\mathrm{eff},i}=n$, but if, two clouds each contain $49$\% of the original cloud's mass, and a third cloud contains the final $2$\%, then $n_{\mathrm{eff},i}\sim2.1$, as intended because here the cloud is primarily divided in two. This prevents a small number of stray particles from being considered as having produced a major merger or separation event.

Having established the methodology behind $n_{\mathrm{eff},i}$ and $f_{\mathrm{r},i}$, we must also ensure that we use robust values based upon ensemble statistics to define global properties that can be compared between simulations. We thus take the arithmetic mean across all clouds $i$ at time $t=t'$ to produce a single $n_\mathrm{eff}$ and $f_\mathrm{r}$ at each point in time for a defined $t'$ for each simulation.

We also calculate the vertical velocity dispersion as a measure of the turbulence in the disc. We choose the vertical velocity dispersion as it excludes the planar components that will have a large contribution from the rotation of the clouds and the disc as a whole. This is determined by calculating
\begin{equation}
\sigma_z = \frac{\sum_i m_i v_{z,i}^2}{\sum_i m},
\end{equation}
across all $N$ gas particles in the simulation. This quantity is not a true particle-particle velocity dispersion, but can be thought of as a measure of the specific energy associated with vertical motions. It is not equivalent to the observed velocity dispersions, partially as our weak feedback produces a low velocity dispersion, but provides a basis for comparison between simulated models.

Additionally, we estimate the recycling timescale by defining
\begin{equation}\label{cyceq}
t_\mathrm{cyc}=\frac{-t}{\ln\left(1-\frac{M_f}{M_g}\right)},
\end{equation}
where $M_f$ is the mass of gas particles that have never been in any cloud, and $M_g$ is the total mass of all gas particles. This definition results from assuming that gas is accumulated on to clouds at a constant rate, and that the mass-fraction of newly accumulated gas that has never been in a cloud before is $M_f/M_g$ -- i.e. the gas is well-mixed. 

\section{Results}\label{resultsect}

\subsection{General evolution}
\begin{figure*}
\includegraphics[width=1.\columnwidth]{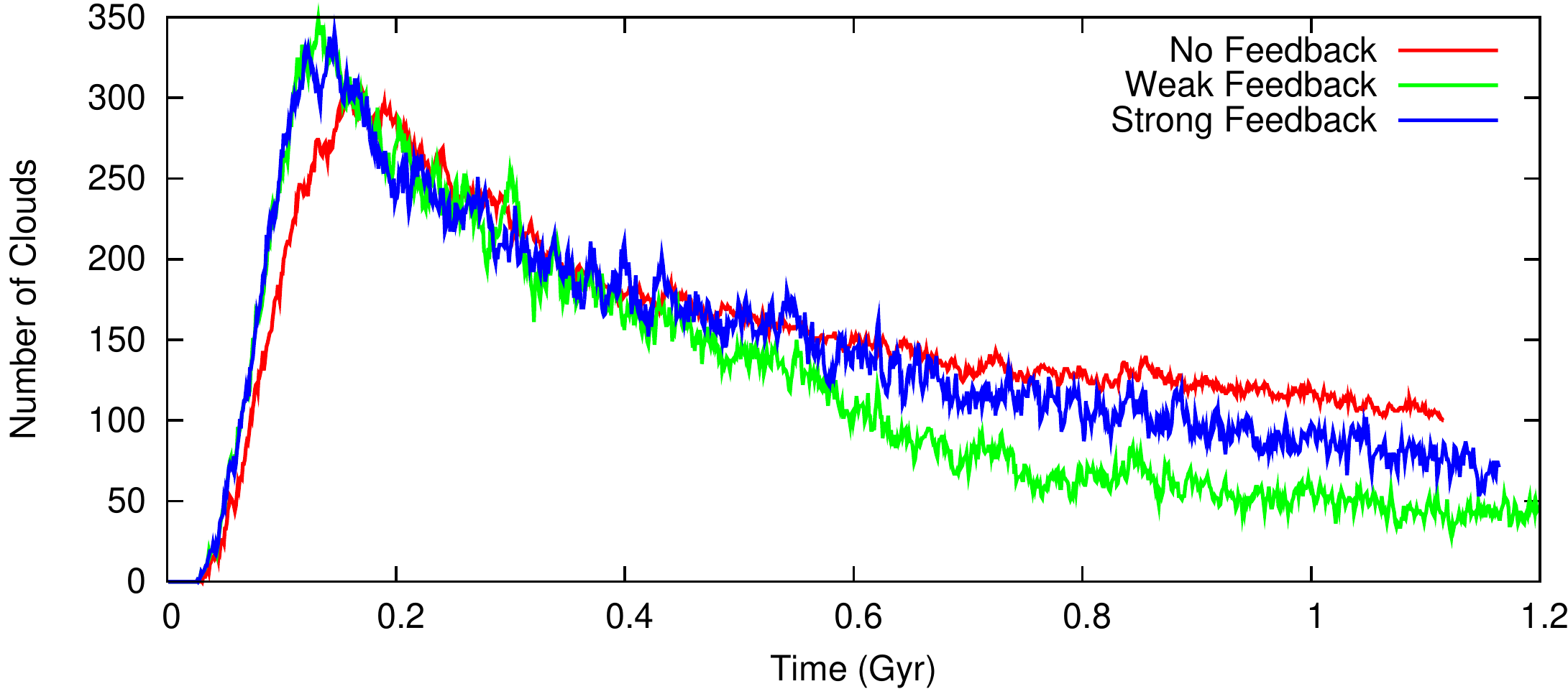}~
\includegraphics[width=1.\columnwidth]{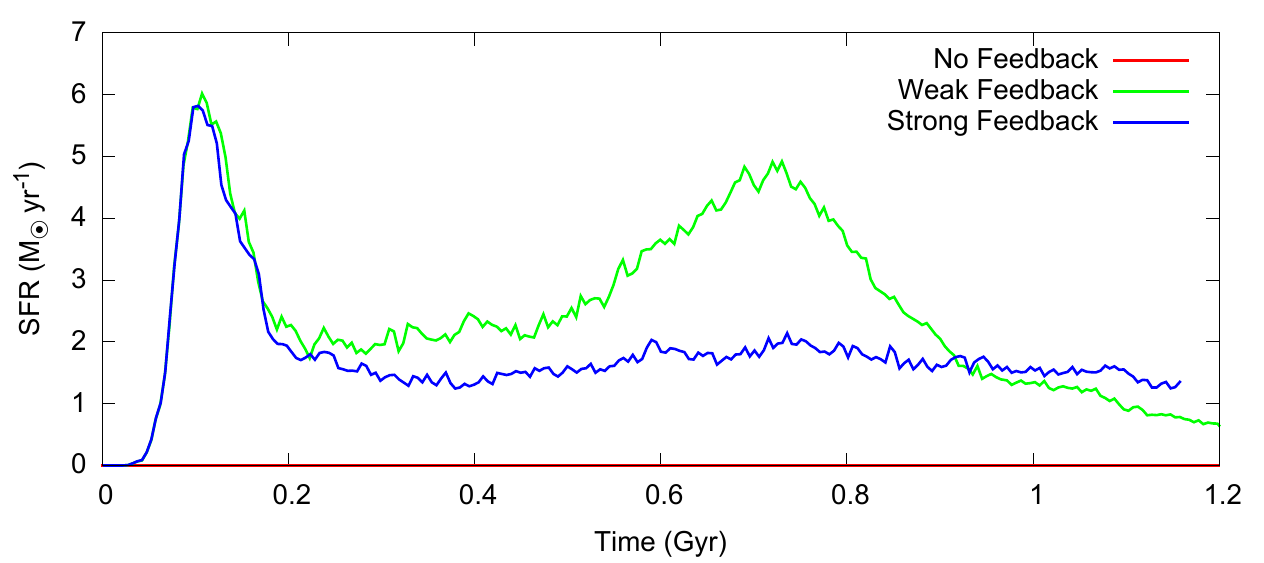}\\
\caption{\label{ncloudssfrtrack} Left: Evolution of the number of clouds in runs with and without feedback. Clouds are identified with an algorithm based on the friends-of-friends method, and then counted. Right: Star-formation rate in runs with and without feedback. The initial peak in both feedback runs coincides with the formation of the first clouds. The second peak in the weak feedback run results from the build-up of massive clouds that are not properly disrupted by the weak feedback. The no feedback run has zero star formation.}
\end{figure*}

\begin{figure*}
\includegraphics[width=1.\columnwidth]{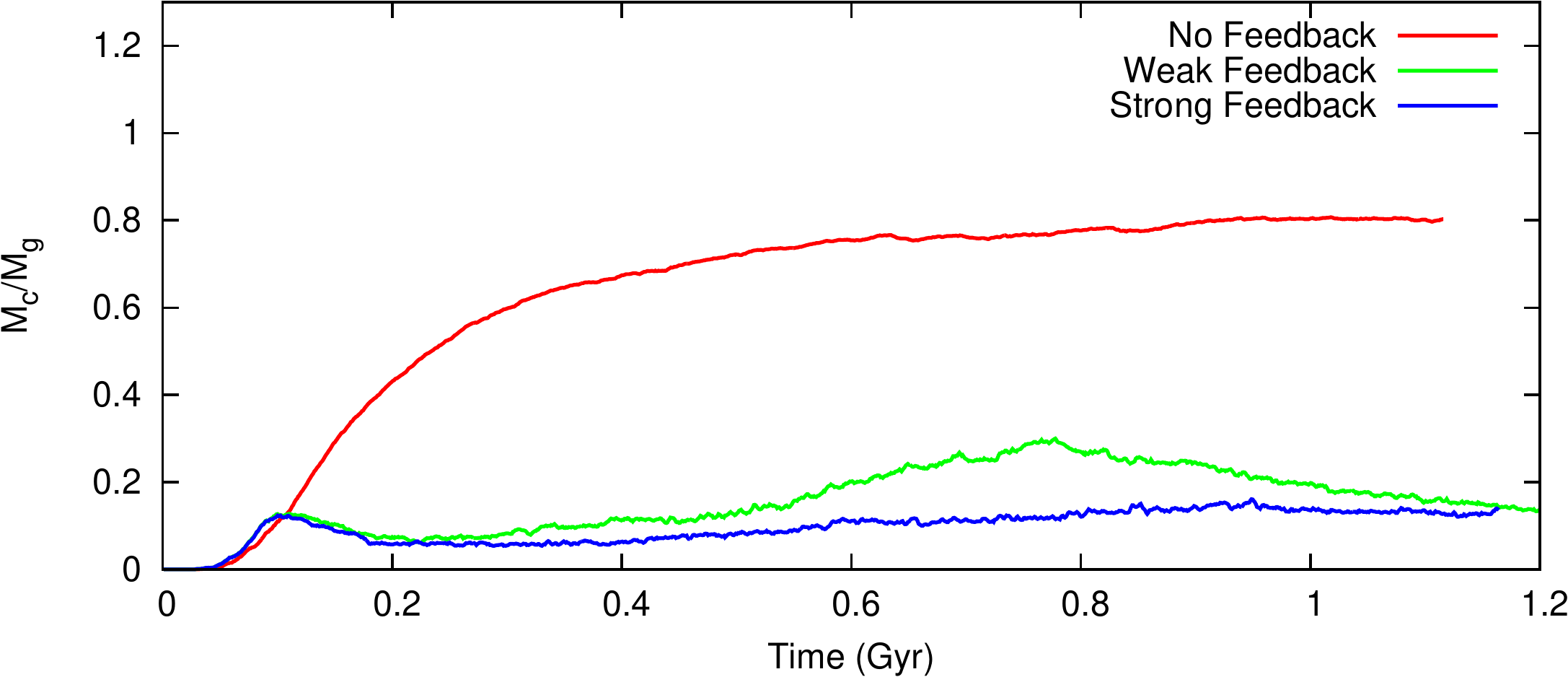}~
\includegraphics[width=1.\columnwidth]{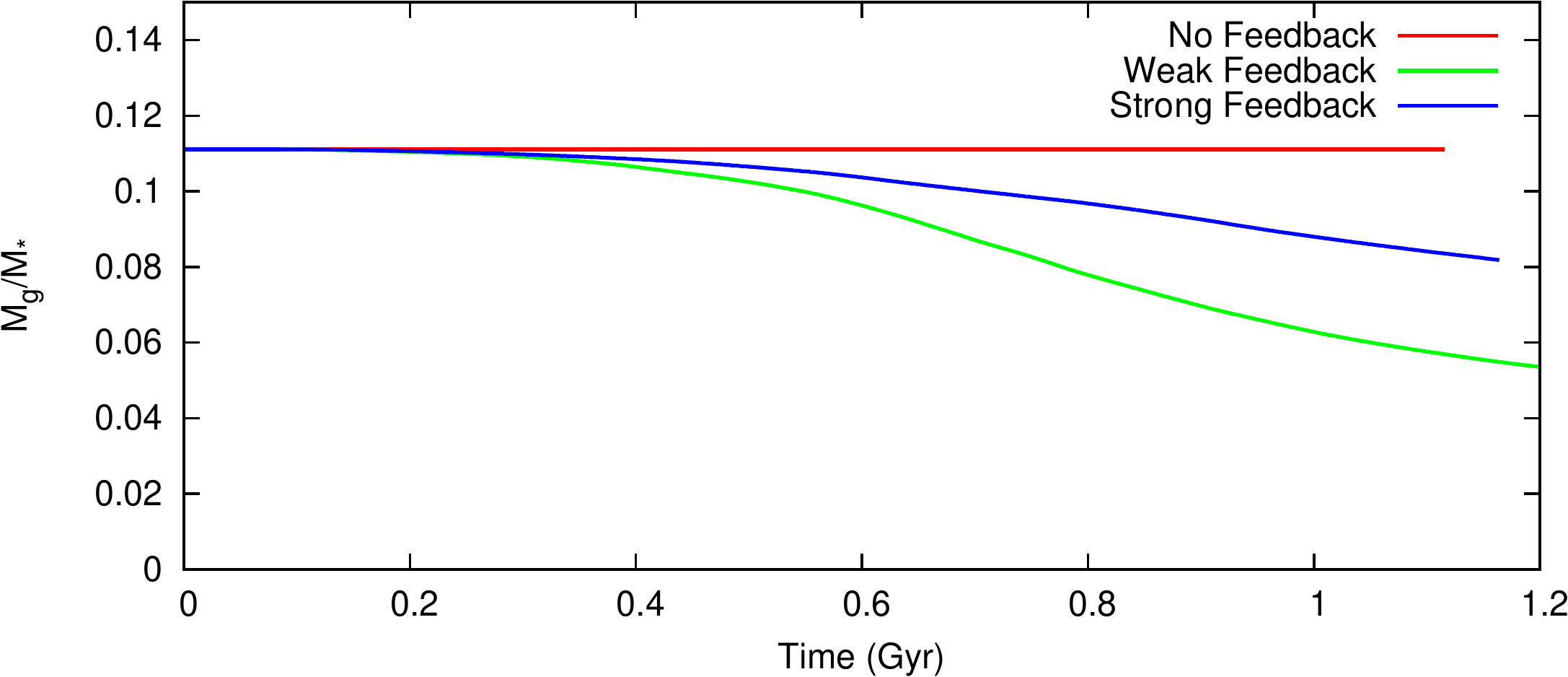}\\
\caption{\label{mcloudsmgas} Left: Evolution of the mass of clouds as a fraction of the total mass of gas particles (i.e. $M_c/M_g$) in runs with and without feedback. In the no feedback model, clouds continually accrete material. With feedback, gas is recycled and less mass is retained within the cloud. The peak at $\sim800$ Myr in the weak feedback run results from the build-up of massive clouds that are not properly disrupted by the weak feedback. Right: Evolution of the ratio of gas-mass to stellar-mass ($M_g/M_*$) in runs with and without feedback. Gas is depleted more rapidly with weak feedback likely because clouds are less disrupted. Gas is not depleted in the no feedback run.}
\end{figure*}

\begin{figure*}
\includegraphics[width=2.2\columnwidth]{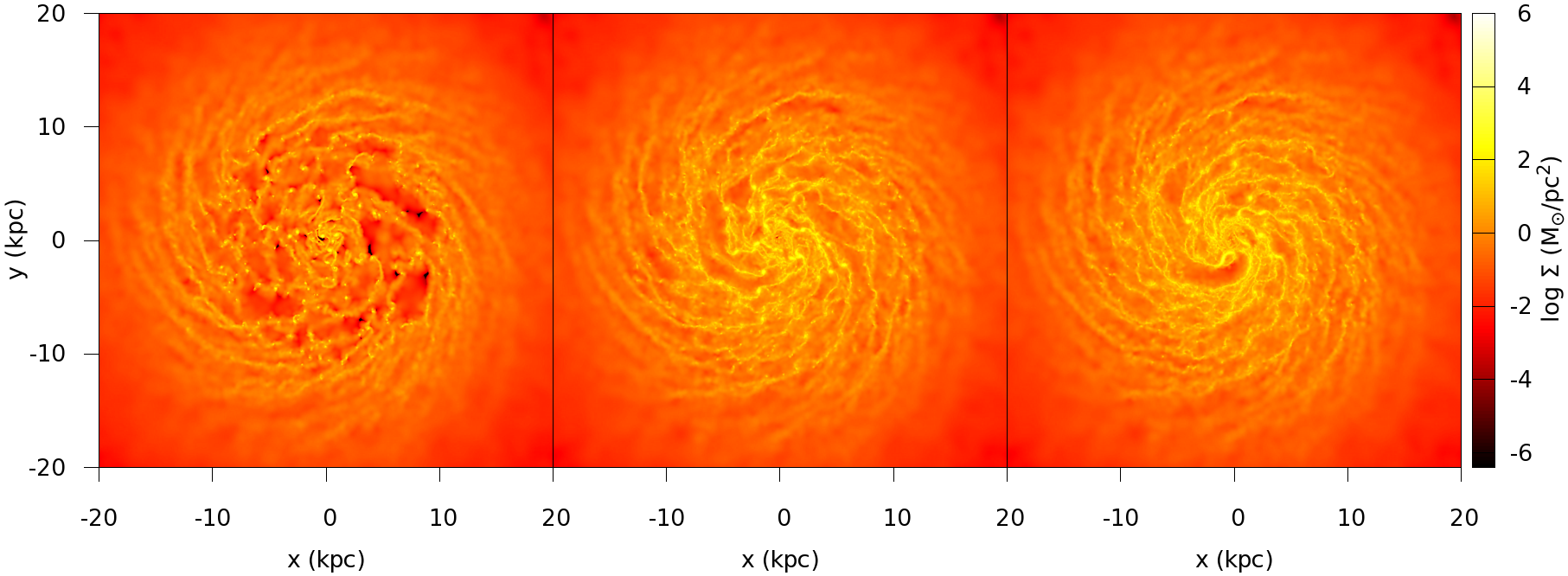}
\caption{\label{surfacedensity} Gas surface density at $t=500$ Myr. Left: No feedback. Centre: Weak feedback. Right: Strong feedback.}
\end{figure*}

\begin{figure}
\includegraphics[width=1.\columnwidth]{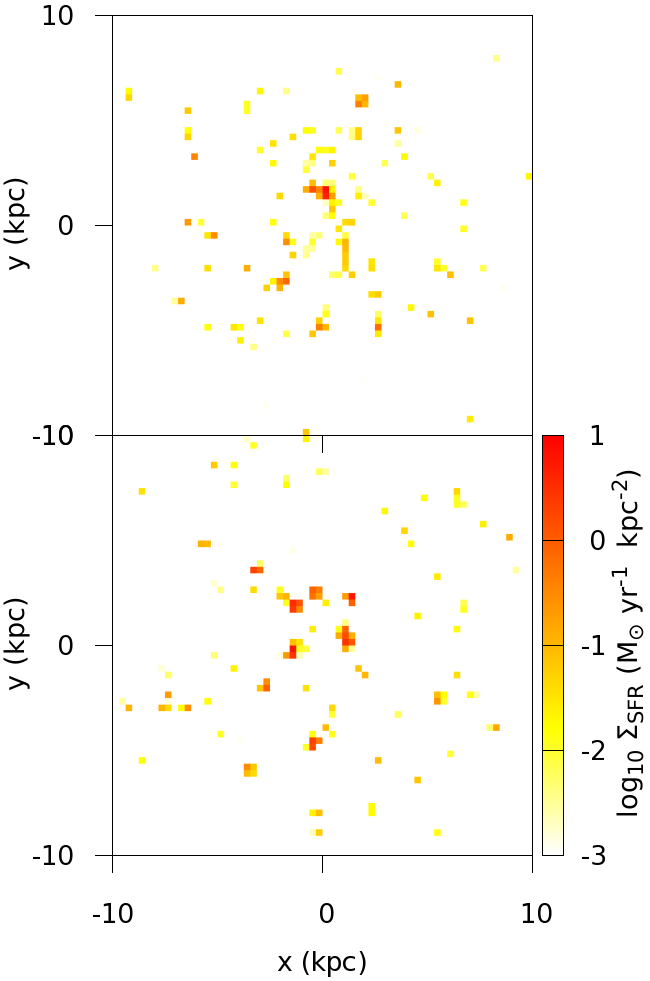}
\caption{\label{sfrdensity} Instantaneous star formation density at $t=700$ Myr. Top: Strong feedback. Bottom: Weak feedback.}
\end{figure}

\begin{figure}
\includegraphics[width=1.\columnwidth]{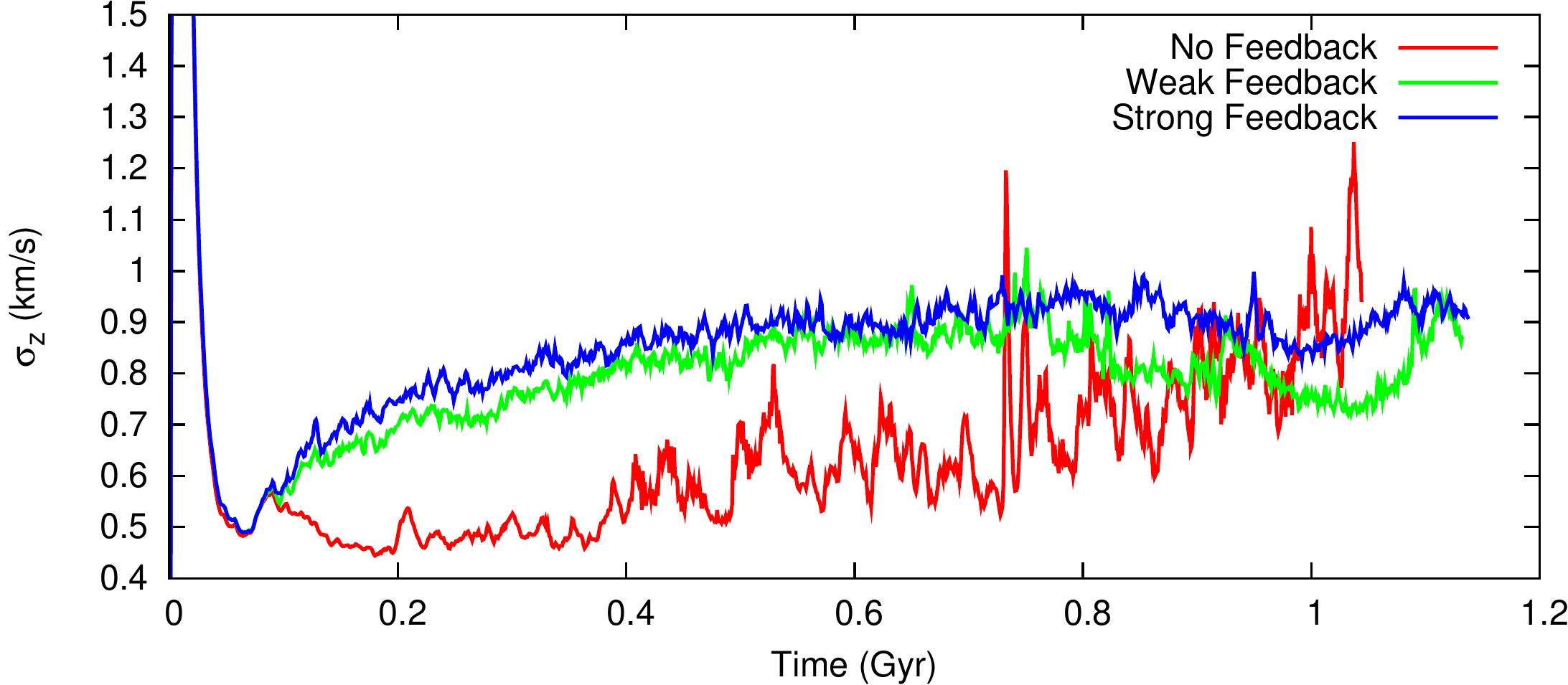}
\caption{\label{gasz} Vertical velocity dispersion, $\sigma_z$ in runs with and without feedback. This is measured across all gas particles in the simulation.}
\end{figure}

\begin{figure}
\includegraphics[width=.5\columnwidth]{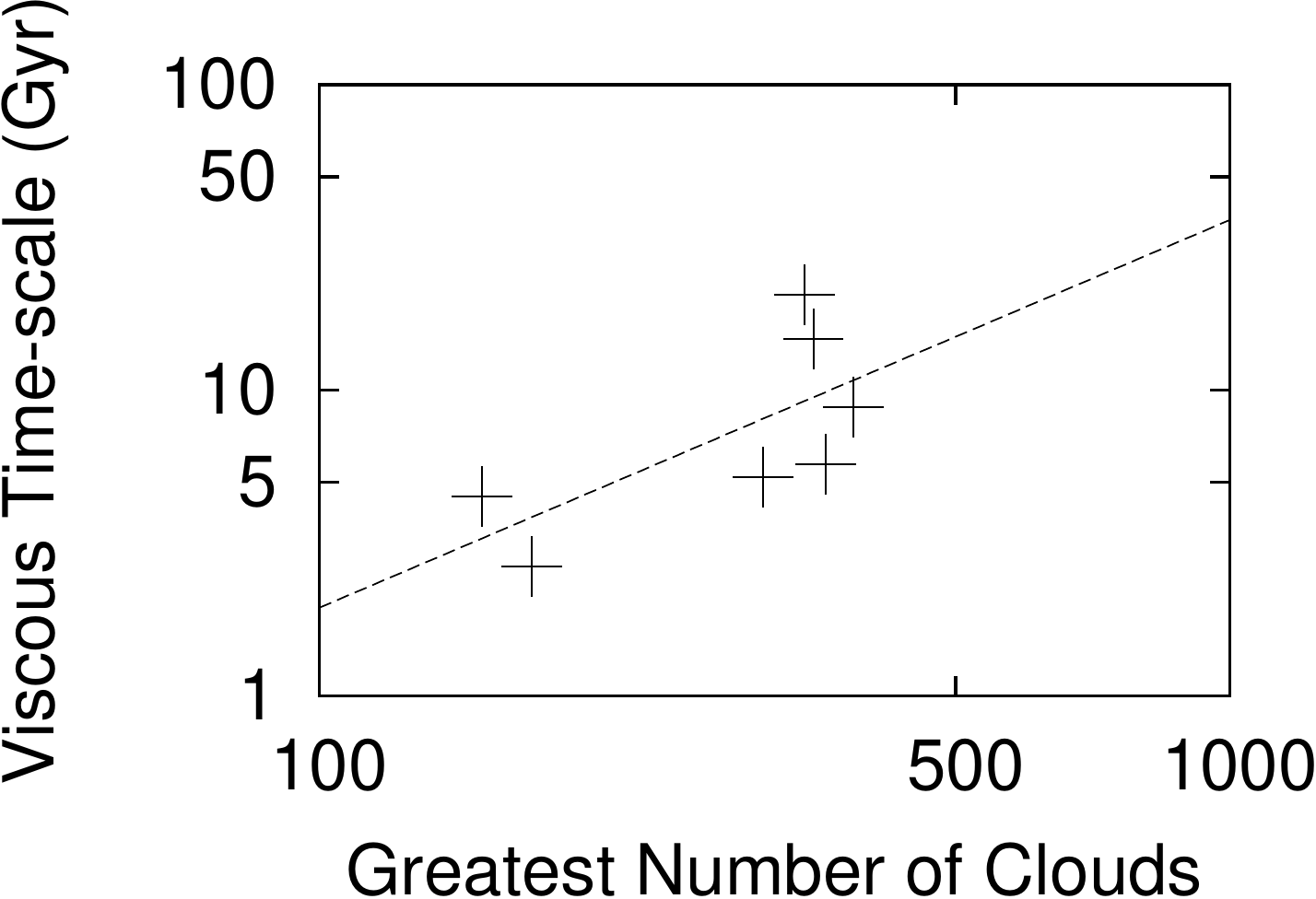}~
\includegraphics[width=.5\columnwidth]{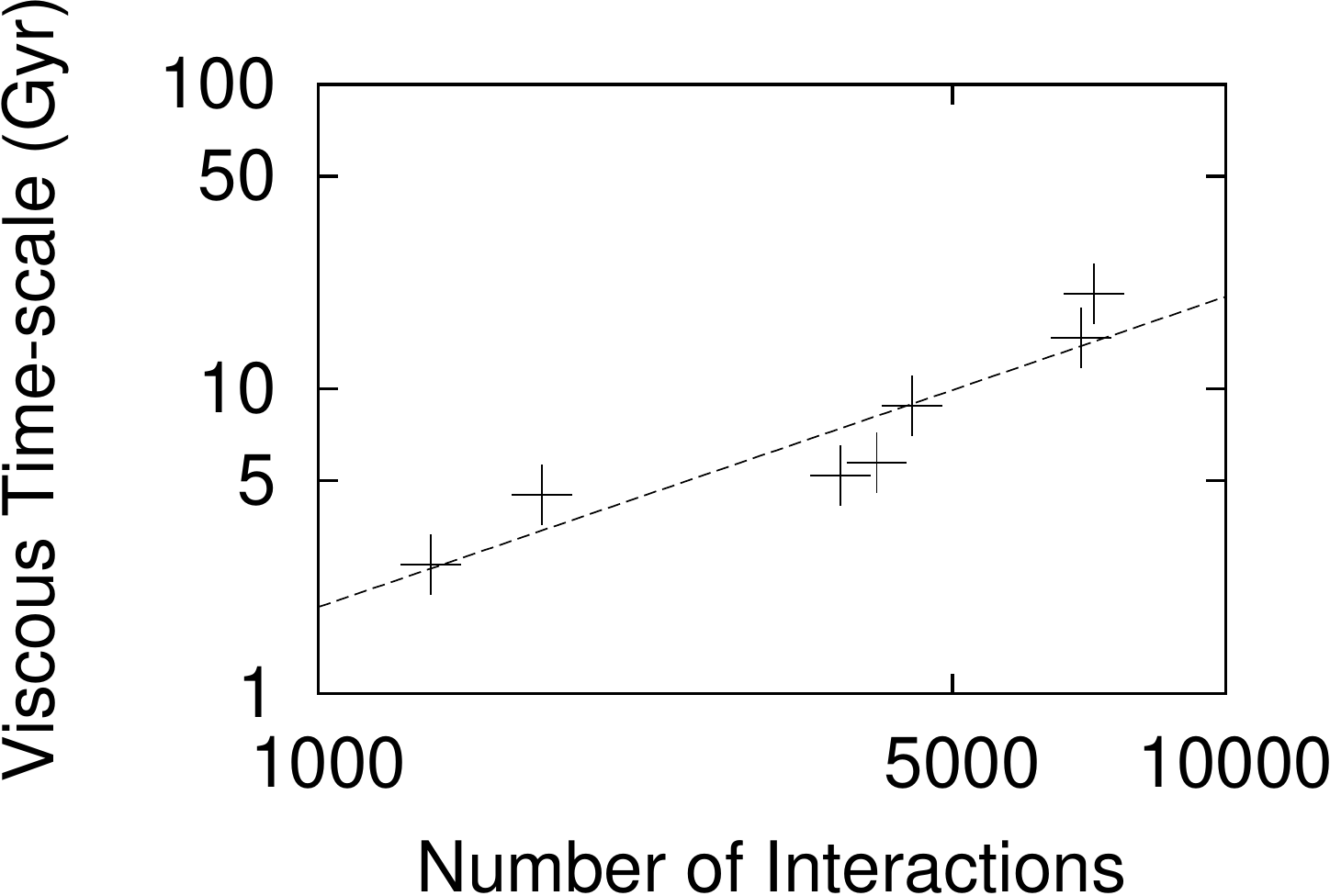}\\
\caption{\label{visctrends} General trends in viscous time-scale across the Milky Way models presented here and in WT12. Least-square power-law fittings are also plotted. Left: Viscous time-scale versus the peak number of clouds in the simulation. Right: Viscous time-scale versus the total number of recorded interactions. These interactions are ``separations'' or ``mergers'' identified by our cloud-tracking algorithm, and do not include longer range interactions.}
\end{figure}


The discs evolved similarly to the Milky Way models such as LowSoftMW in WT12. The onset of cooling at the beginning of the simulation causes the disc to flatten, and the gaseous disc then becomes Toomre unstable, and fragments into a large number of small clouds (Fig.~\ref{surfacedensity}). As the simulation evolves, these clouds orbit, and collide with each other. Initially a large number of clouds are produced, but this number decreases as clouds merge (Fig.~\ref{ncloudssfrtrack}). The lack of a dynamic temperature floor causes the simulations without feedback to form clouds slightly earlier, as star formation and feedback has not yet occurred (Fig.~\ref{ncloudssfrtrack}), and so there is no restraint to gas cooling.

Without feedback, clouds are both dense and comparatively strongly bound, and are hence not easily disrupted (as we show in Section~\ref{fraghist}), and so the number of clouds decays more slowly than in the simulations with feedback. The strong feedback run produces somewhat more numerous (and less massive) clouds than the weak feedback run, but it might be considered surprising that the no feedback cloud produces an even greater number of clouds. However, it is clear in Fig.~\ref{mcloudsmgas} that the nature of these clouds is very distinct: in the no feedback run there is considerably more mass in clouds than in both the weak and strong feedback models. With feedback, the total mass in clouds decreases after the initial peak, but in the no feedback run, clouds are not easily disrupted and continue to accrete gas. The result is that the no feedback run contains a large number of massive clouds, as these clouds have very long life-times.


Weak feedback is less efficient at disrupting dense star-forming regions, allowing clouds to grow more numerous and more massive. Indeed, a small number of clouds are produced that are too massive to be disrupted by the weak feedback. These clouds are massive and have very high star formation rates (Fig.~\ref{sfrdensity}), resulting in a peak in the cloud mass and star formation rates at $t\approx700$ Myr, until the clouds eventually lose enough mass to dissipate. This agrees with the simulations of \citet{2011MNRAS.417.1318D}, where only the weakest feedback allowed very large clouds to form. Even prior to this peak, the more massive and more numerous clouds in the weak feedback run generate a higher star formation rate than in the strong feedback model. This star formation consumes gas rapidly (Fig.~\ref{mcloudsmgas}, right), eventually quenching molecular both cloud formation formation and star formation (Fig.~\ref{ncloudssfrtrack}, right) in the weak feedback model, and so at later times ($t\approx1.2$ Gyr), the mass-fraction of gas in clouds is similar in both feedback simulations (Fig.~\ref{mcloudsmgas}, left).

The vertical velocity dispersion, $\sigma_z$ is plotted in Fig.~\ref{gasz}. Overall the velocity dispersions are low, as a result of our overall weak feedback. The evolution in all runs is almost identical until the initial peak in star and cloud formation. At this point, the thermal feedback causes the feedback runs to increase their velocity dispersion until reaching an equilibrium of $\sigma_z\approx0.9$ km s$^{-1}$ at $t\approx500$ Myr. However, the no-feedback run has a smaller velocity dispersion ($\sigma_z\approx0.5$ km s$^{-1}$) that gradually increases over the course of the simulation. This is likely because scattering events between clouds are the only way to build a vertical velocity dispersion without feedback, and as these clouds are long-lasted, this velocity dispersion accumulates slowly.

\subsection{Viscous time-scales}

\begin{table}
\begin{tabular}{ cl | c | c | c | c | c |}
\hline\hline
Name & $N_{int}$ & $t_\nu$ (Gyr)& $f_\mathrm{pro}$ & $f_{<30^\circ}$\\
\hline
No Feedback & 3635 & 5.3 & 0.88 & 0.63\\
Weak Feedback & 6932 & 14.7& 0.84 & 0.43\\
Strong Feedback & 7159 & 20.5 & 0.82 & 0.34\\
\hline
\end{tabular}
\caption{\label{visctimes}Summary of simulation results. The viscous time-scales are the mean time-scales for the first $1$ Gyr of simulation. For each simulation we also show the fraction of prograde clouds $f_\mathrm{pro}$, and the fraction of strongly prograde clouds $f_{<30^\circ}$ whose angular momentum axies are less than $30^\circ$ from that of the galaxy. These prograde fractions at taken at $t=500$ Myr.} 
\end{table}

\begin{figure}
\includegraphics[width=1.\columnwidth]{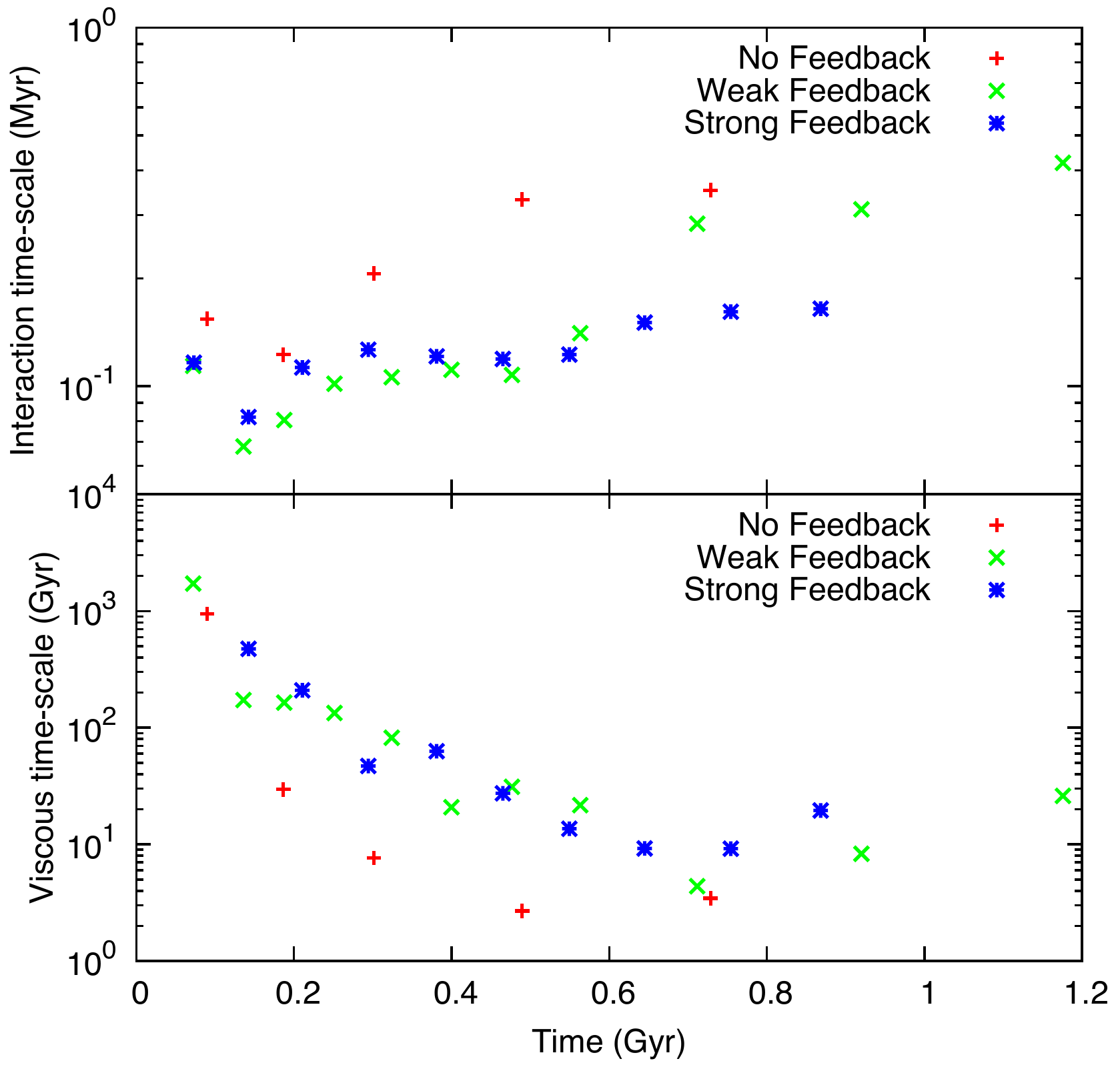}\\
\caption{\label{viscevolve} Evolution of the viscous time-scale and interaction time-scale for the three simulations presented in this paper. Each point represents the effect of $700$ interactions between clouds.}
\end{figure}

The viscous time-scales in models with feedback are longer than in models without feedback, and the strong feedback model has a longer viscous time-scale than the weak feedback model (Table~\ref{visctimes}). There is a weak trend where the viscous time-scale increases with the peak number of clouds formed in the simulation, but this is likely a side-effect of a stronger trend, namely that the viscous time-scale depends on the number of recorded interactions (mergers and separations) in the simulation (Fig.~\ref{visctrends}). A power law fitting gives a power index of $1.0\pm0.2$, consistent with a linear relationship. Defining $t_c=1/R_c$ as the collisional time-scale (where $R_c$ is the rate of merger \& separation events, which is proportional to the total number of these interactions in a fixed time interval), and $\eta$ as the mean fraction of a cloud's kinetic energy lost in a collision, the viscous time-scale is equal to $t_\nu=t_c/\eta$. Our results thus imply $\eta\propto t_c^2$.
We can interpret this result based on WT12 and earlier work \citep{1978Icar...34..227G,2002ApJ...581.1013B}. If collisions are common, then $\eta\propto v_s^2 \propto 1/t_c^2$, where $v_s$ is the velocity dispersion. But if collisions are rare, then $\eta\propto \Delta R^2\propto t_c^2$, where $\Delta R$ is the radial excursion of a cloud in its orbit. Our result shows that the rare collision case is more applicable to a population of molecular clouds in a Milky Way-type galaxy.
However, this correlation is only clear when contrasting between the viscous time-scale calculated from the cumulative effect of all interactions (mergers and separations) detected in the first $1$ Gyr of a simulation. If we instead calculate the viscous and interaction time-scales from a fixed number of interactions that are consecutive in time, we can find the time evolution of the viscous and interaction time-scales. Using $700$ interactions per point (Fig.~\ref{viscevolve}), we find that the relationship between $t_\nu$ and $t_c$ is not clear, and conclude that the details of the cloud population and their interactions are more significant on this shorter time-scale.

\subsection{Mass spectra}

\begin{figure*}
\includegraphics[width=1.\columnwidth]{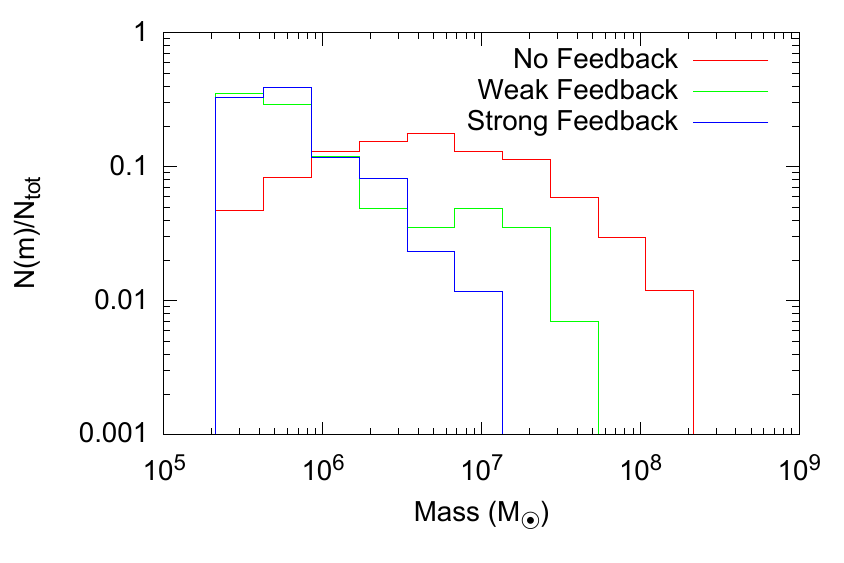}~
\includegraphics[width=1.\columnwidth]{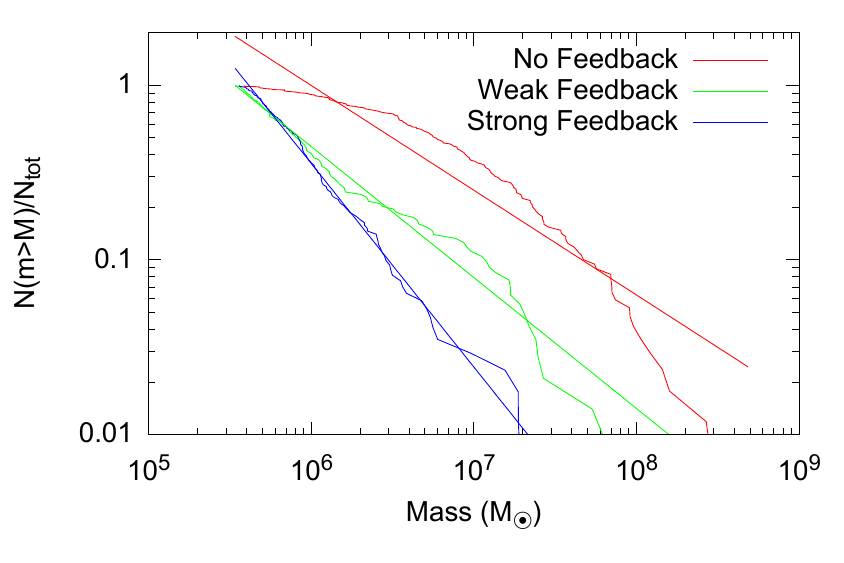}\\
\caption{\label{massspect} Left: Histograms of cloud masses at $t=500$ Myr. Feedback disrupts the most massive clouds, while encouraging the formation of small clouds. Right: Cloud cumulative mass spectra at $t=500$ Myr, with power-law fittings to $N(m>M)\propto m^{\alpha+1}$.}
\end{figure*}

\begin{figure*}
\includegraphics[width=.67\columnwidth]{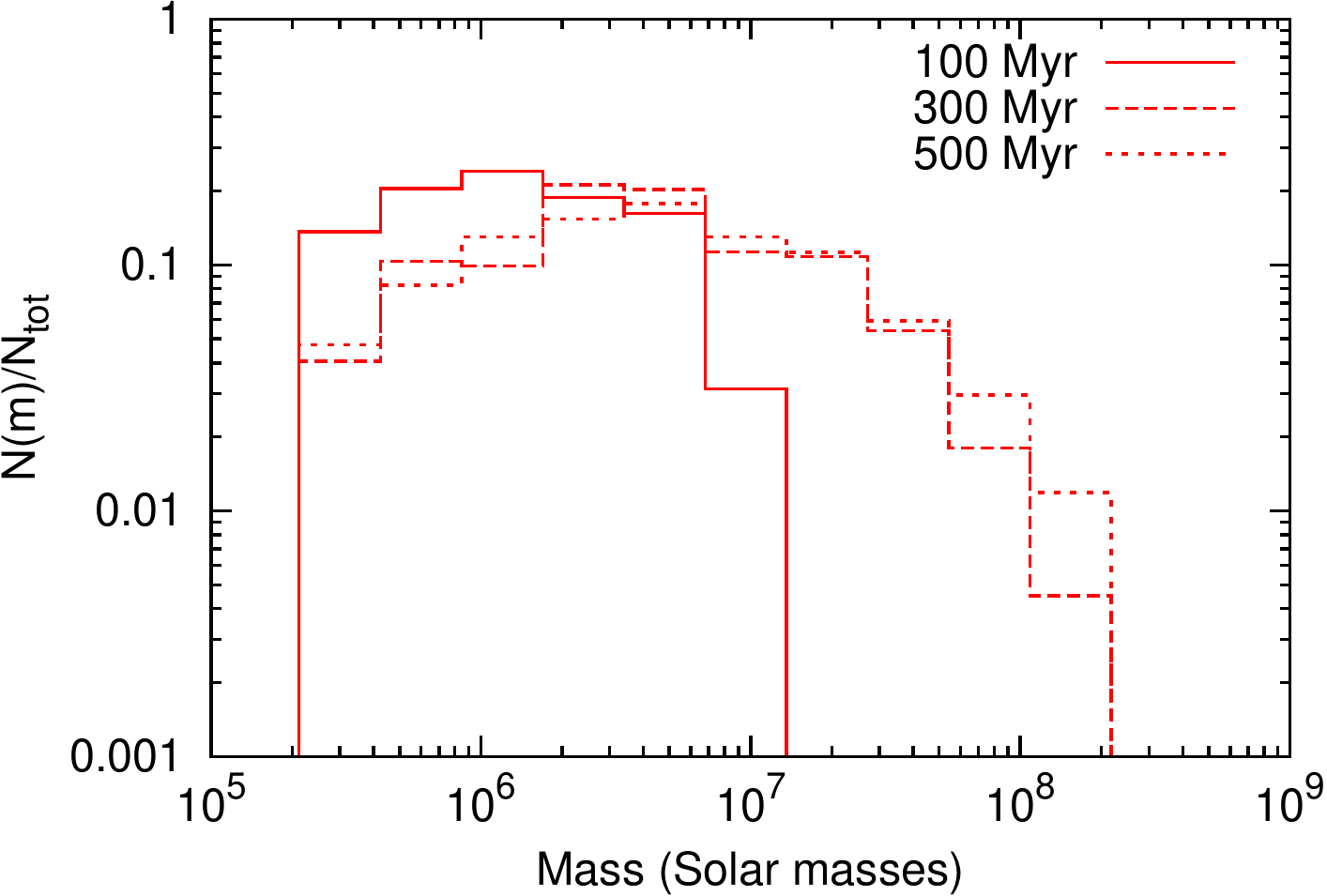}~
\includegraphics[width=.67\columnwidth]{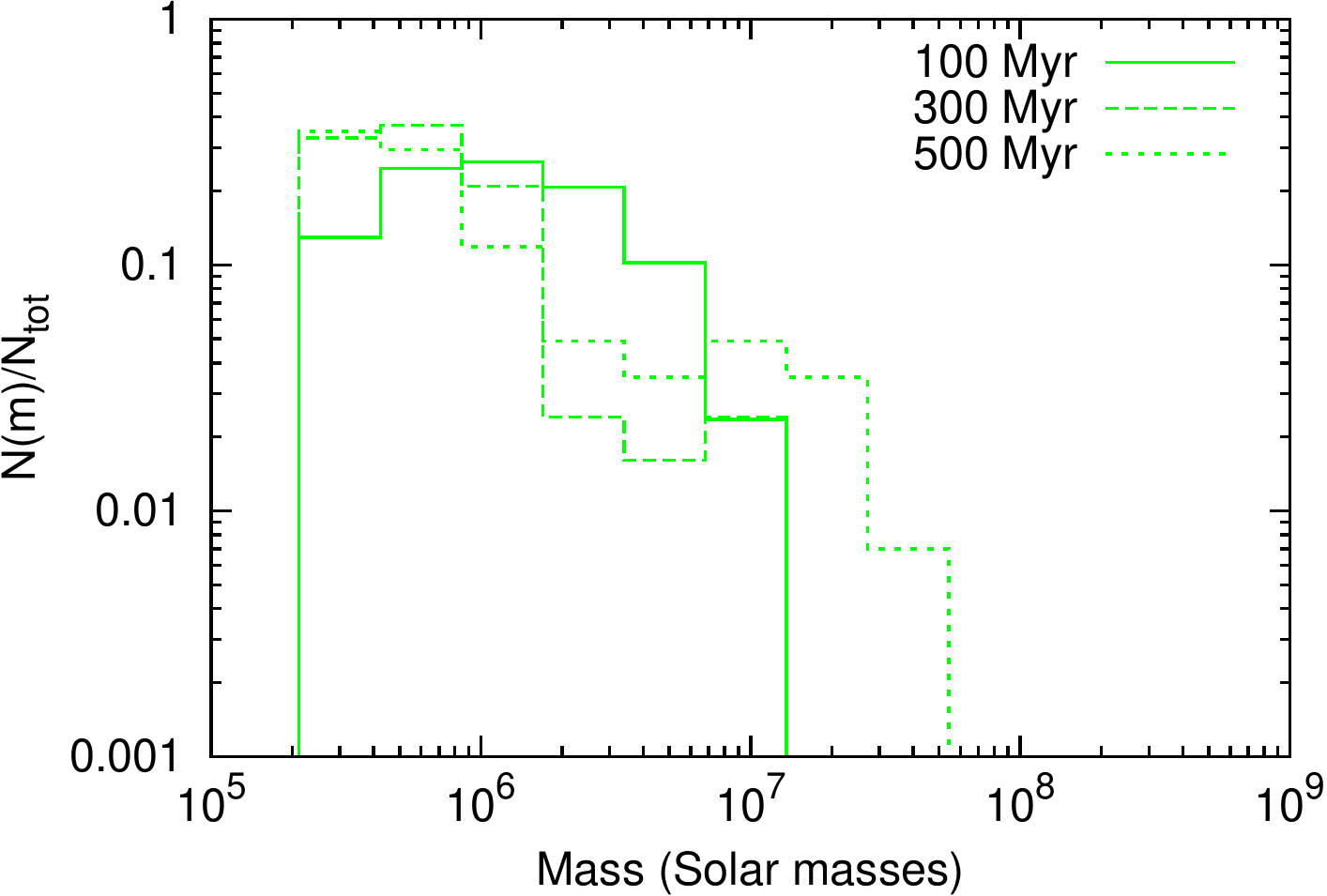}~
\includegraphics[width=.67\columnwidth]{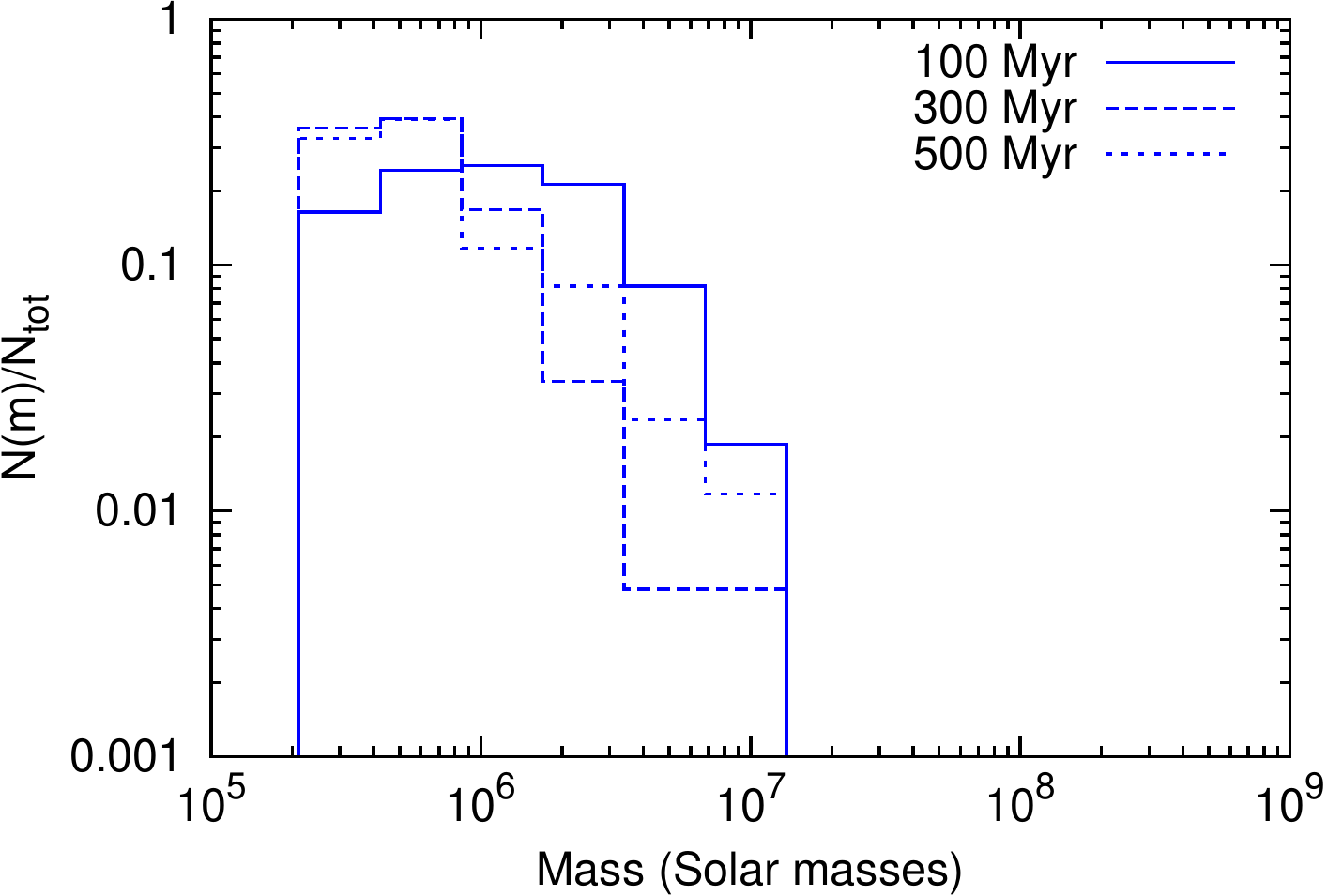}\\
\caption{\label{evolvespect} Evolution of cloud masses from $t=100$ Myr to $t=500$ Myr. Left: No feedback. Centre: Weak feedback. Right: Strong feedback. The most massive clouds do not form immediately, and only when feedback is not present or weak. Feedback only starts to take effect after $100$ Myr, producing a larger number of small clouds.}
\end{figure*}

Mass spectra for all three simulations at $t=500$ Myr are plotted in Fig.~\ref{massspect}. Feedback disrupts clouds, and so the clouds are less massive on average as feedback increases in strength. When feedback is included, the number of clouds in each bin decreases almost monotonically with mass, probably because with stronger feedback it becomes increasingly unlikely that a cloud will survive long enough to accrete enough material or undergo enough mergers to reach a large mass. Without feedback, clouds are free to merge hierarchically, and less gas is redistributed to the diffuse ISM, quenching the formation of new small clouds, which may explain the paucity of low-mass clouds in the simulation without feedback. The evolution of cloud masses is shown in Fig.~\ref{evolvespect}. All three simulations have similar mass spectra at $t=100$ Myr. At later times in the no feedback simulation, low mass clouds are depleted as clouds merge, producing a population of high mass clouds, similar to the high mass tail of the no feedback simulations of \citet{taskertan}. In the strong feedback run, a larger number of low mass clouds are produced at later times as feedback causes clouds to be formed from a hotter, more turbulent ISM. The high mass tail is lost, as when diffuse feedback was included in \citet{2009ApJ...700..358T}. In the weak feedback run, both the high-mass and low-mass clouds are produced.

The cumulative mass spectra roughly follow power laws (i.e. $dN/dM\propto m^\alpha$ or $N(m>M)\propto M^{\alpha+1}$). Least-squares fitting gives $\alpha=-1.60\pm0.02$ without feedback, $\alpha=-1.75\pm0.01$ with weak feedback, and $\alpha=-2.166\pm0.009$ with strong feedback. The simulation with weak feedback fits the values of $-1.5$ to $-1.6$ from observations \citep{1985ApJ...289..373S,1987ApJ...319..730S,1989ApJ...339..919S,1997ApJ...476..166W,2010ApJ...723..492R}.

\subsection{Cloud fragmentation histories}\label{fraghist}

\begin{figure}
\includegraphics[width=.98\columnwidth]{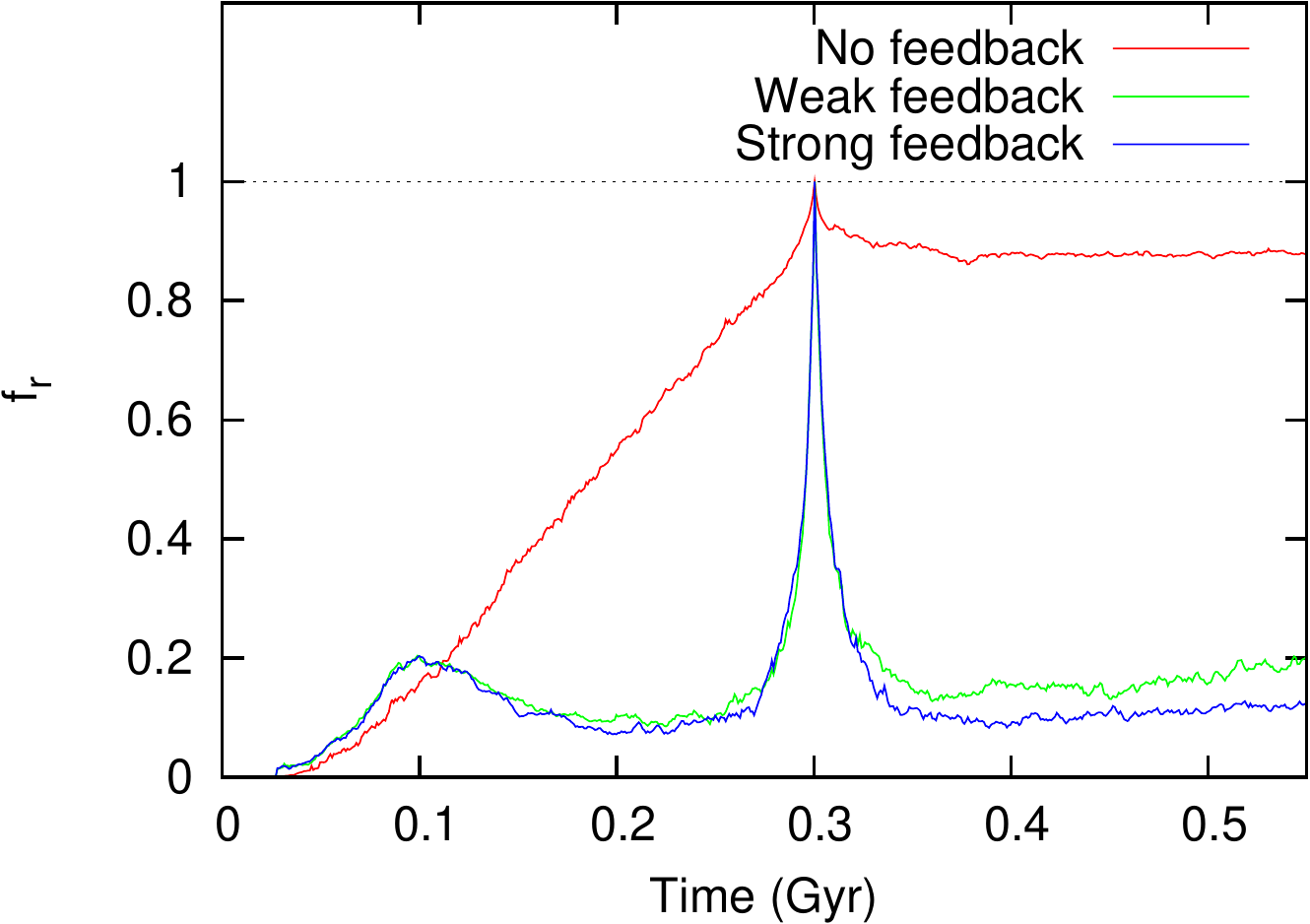}\\
\includegraphics[width=.98\columnwidth]{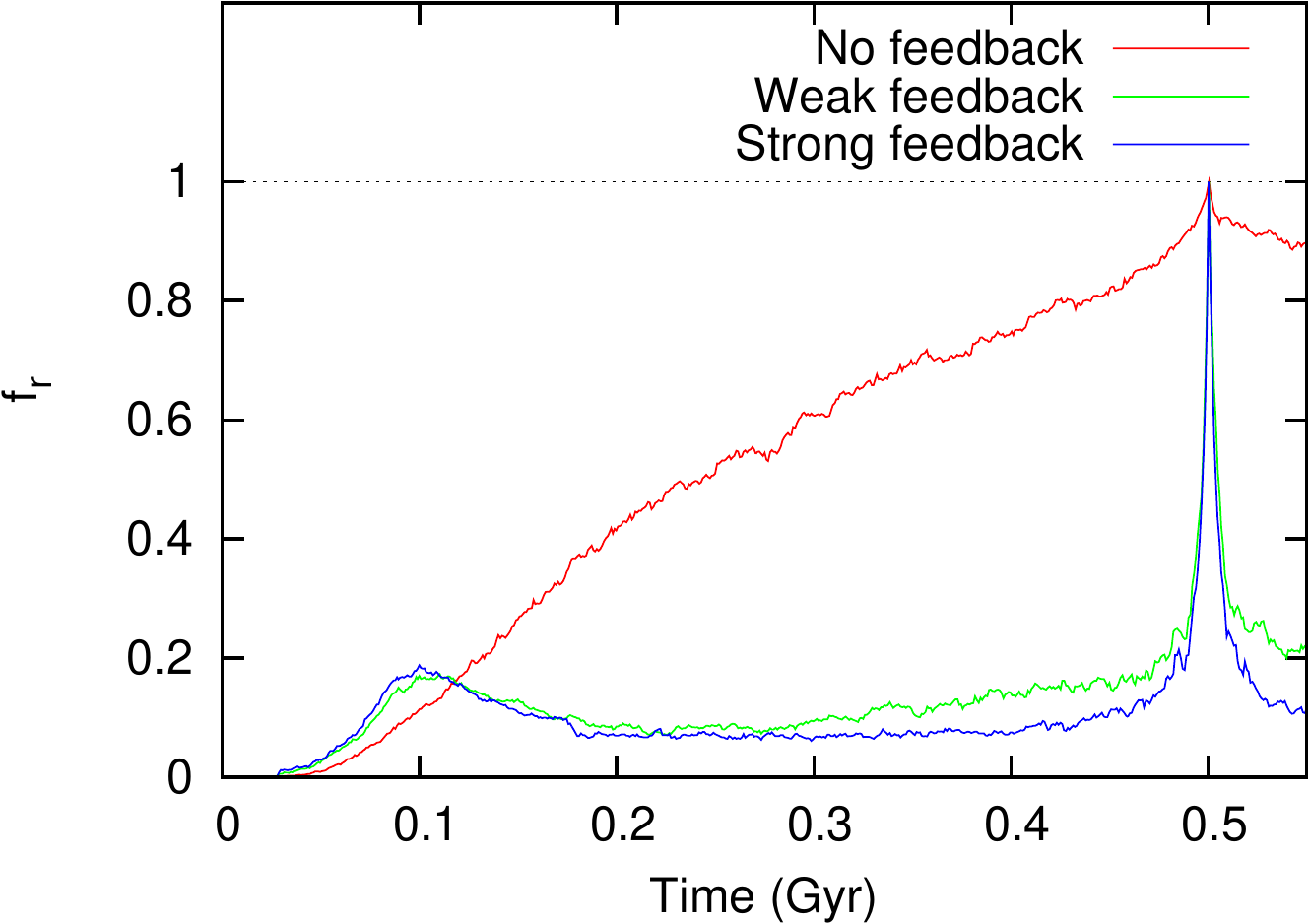}\\
\caption{\label{cloudneff} $f_\mathrm{r}$, the fraction of particles from clouds at $t=t'$ that remain in any cloud. Top: $t'=300$ Myr. Bottom: $t'=500$ Myr. With feedback, gas is only briefly retained in clouds. Without feedback, gas slowly accumulates in a cloud and remains within that cloud.}
\end{figure}

\begin{figure}
\includegraphics[width=.98\columnwidth]{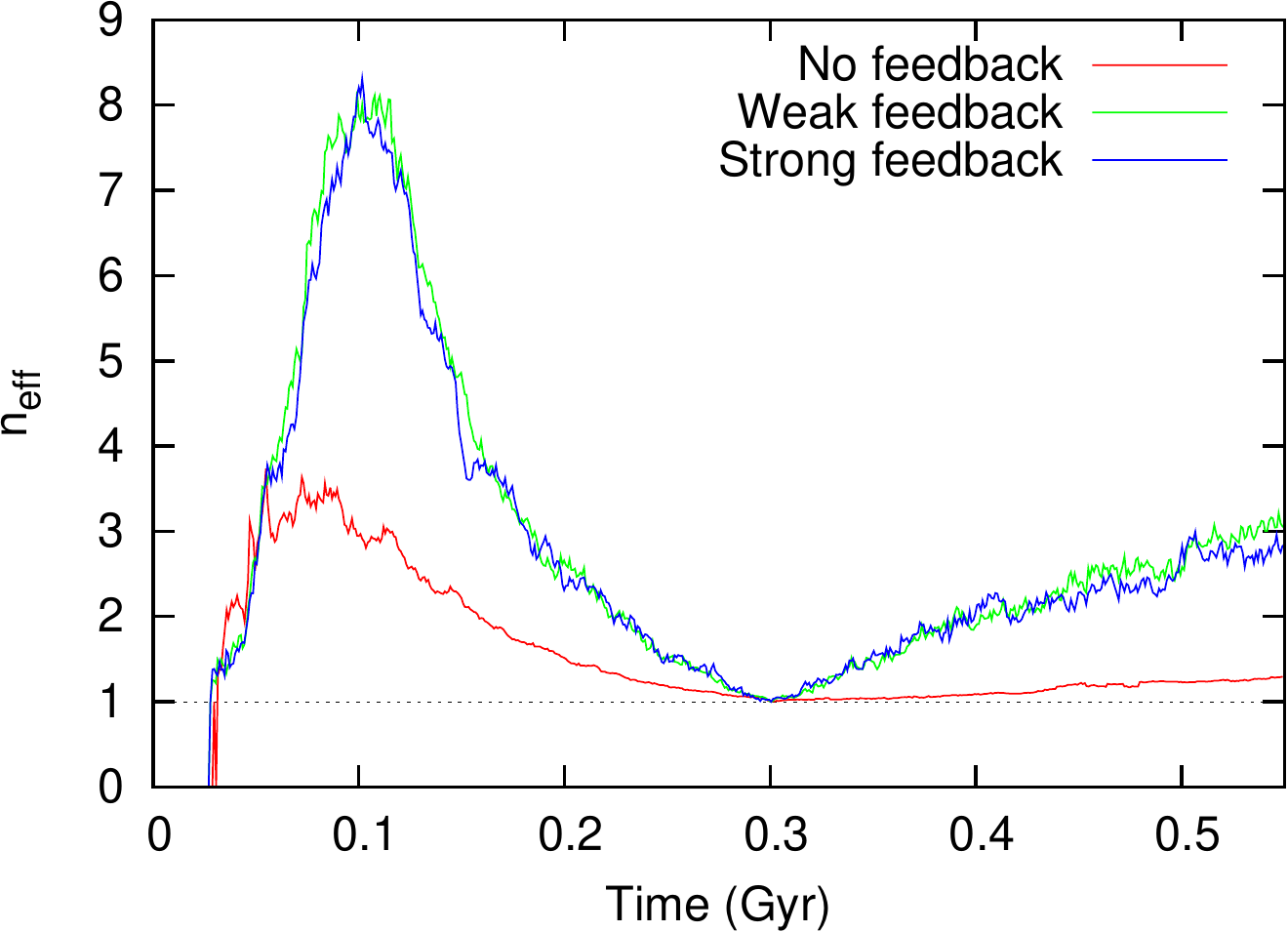}\\
\includegraphics[width=.98\columnwidth]{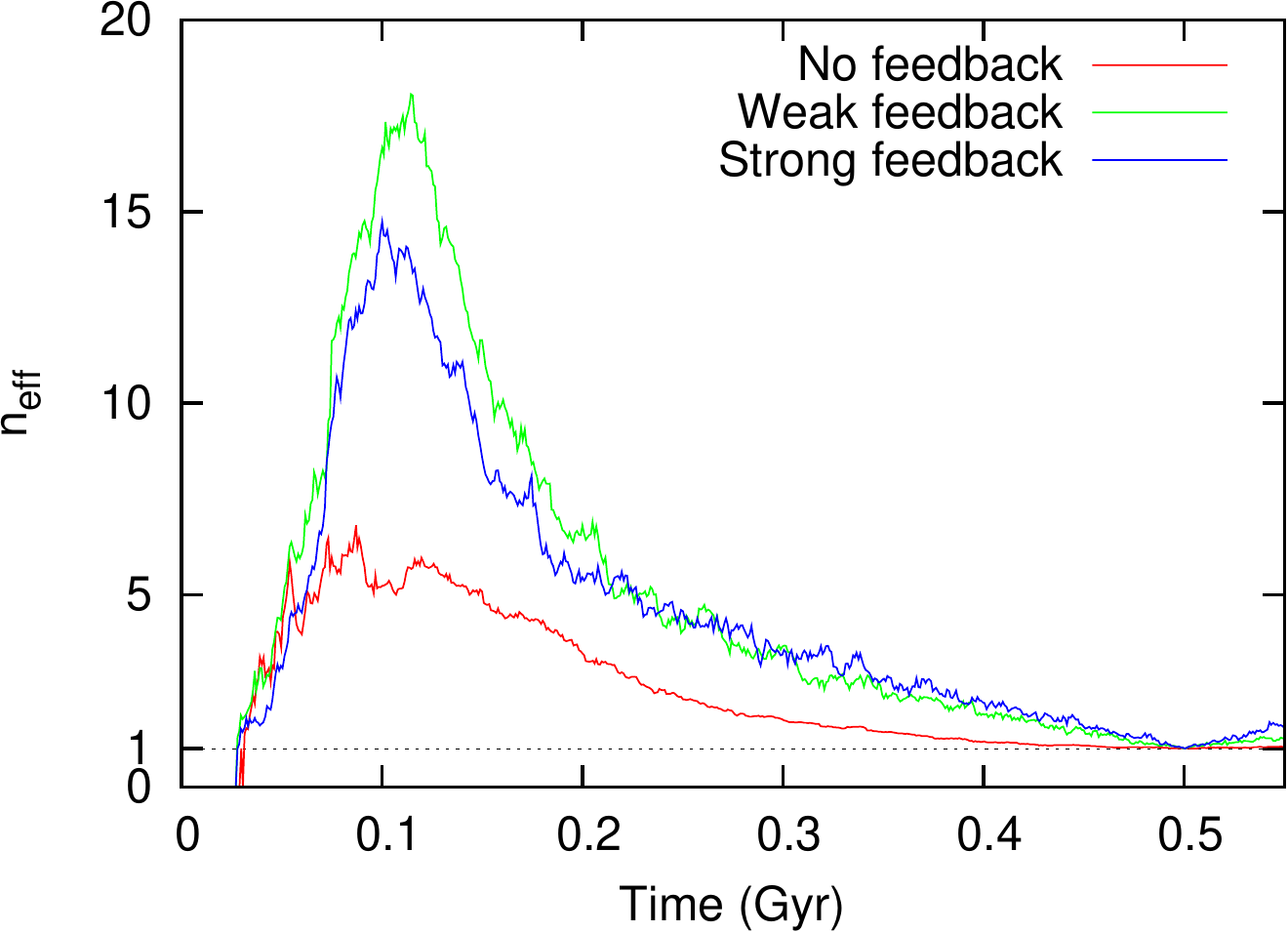}\\
\caption{\label{clouddistf} $n_\mathrm{eff}$, the average of the effective number of clouds the particles from clouds at $t=t'$ have fragmented into over time. Top: $t'=300$ Myr. Bottom: $t'=500$ Myr. Feedback increases $n_\mathrm{eff}$, suggesting a more violent, fragmentary cloud population.}
\end{figure}

In Figs.~\ref{cloudneff} and \ref{clouddistf} we plot the evolution of $f_\mathrm{r}$ and $n_\mathrm{eff}$ for $t'=300$ Myr and $t'=500$ Myr. Note that at $t=t'$, it is necessary for $f_\mathrm{r}$=$n_\mathrm{eff}$=1.

In the no-feedback model, $f_\mathrm{r}$ gradually increases to this point, and then maintains $f_\mathrm{r}\sim0.9$ for $t>t'$, showing explicitly that cold gas is not being recycled back into the ISM. It also indicates that these clouds did not form recently from the diffuse ISM, but instead either formed from conglomerations of smaller clouds, or as isolated clouds that continually accrete from the diffuse ISM.

The plots of $n_\mathrm{eff}$ indicate that in the no feedback run, material from clouds at $t=t'$ were in either $\sim3$ or $\sim5$ clouds at $t=100$ Myr, suggesting that some conglomeration occurs. For $t>t'$, we find that $n_\mathrm{eff}\sim1$. Hence the overall picture in the model without feedback is that clouds form primarily through a small number of hierarchical mergers and remain intact: shear forces and violent interactions between clouds are insufficient to disrupt the clouds.

In the simulations with feedback, we see different behaviour. There is a sharp peak in $f_\mathrm{r}$ at $t=t'$, showing that clouds largely form from the diffuse ISM, and are rapidly dispersed back into diffuse gas. Only a small fraction ($f_\mathrm{r}\sim0.1$) of the gas remains in clouds before or after $t=t'$. This small quantity of gas is spread amongst a larger number of clouds (a larger $n_\mathrm{eff}$) than in the simulations with feedback. This suggests, perhaps expectedly, that gas recycling is more efficient in simulations with feedback.
\begin{figure}
\includegraphics[width=1.\columnwidth]{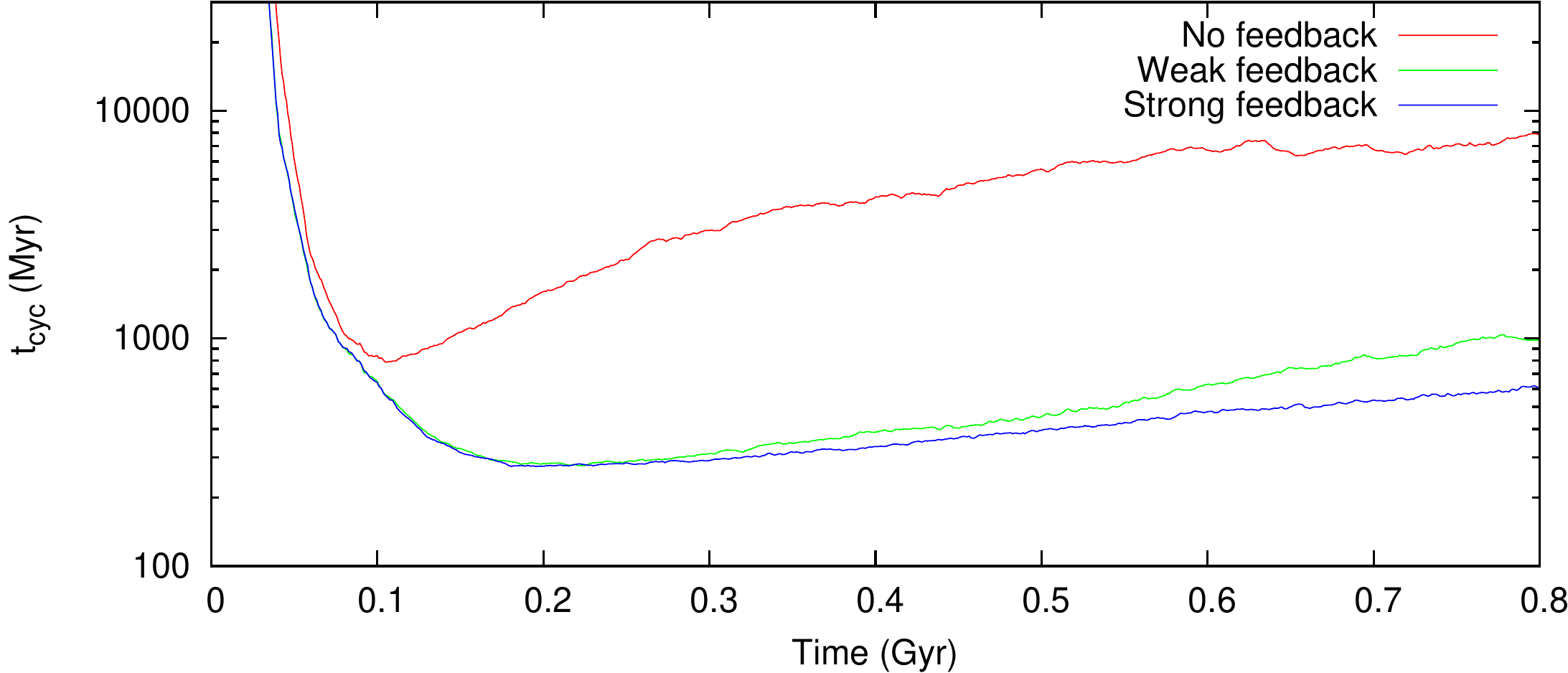}\\
\caption{\label{tcyc} Gas recycling times for all simulations, derived from equation~\ref{cyceq}. }
\end{figure}

The recycling timescales are plotted in Fig.~\ref{tcyc}. In all runs the recycling timescale rapidly drops as clouds initially form at $t\approx100$ Myr. In the feedback runs, these reach a minimum of $t_\mathrm{cyc}\approx300$ Myr, and then slowly increase as gas is depleted, but still remaining below $1$ Gyr. However, in the no feedback run, $t_\mathrm{cyc}$ only just drops below $1$ Gyr before increasing to near $10$ Gyr: unsurprisingly, recycling is not efficient in the absence of feedback. Hence the overall picture in the models with feedback is that clouds form directly from the diffuse and largely recycled ISM, and that clouds are efficiently dispersed back into the diffuse ISM and mixed with the gas from other clouds.

\subsection{Cloud angular momentum}

\begin{table*}
\begin{tabular}{ccccccccc}
\hline\hline
Name & $N_\mathrm{pro}$ & pro$\rightarrow$pro & ret$\rightarrow$pro & diff$\rightarrow$pro & $N_\mathrm{ret}$ & ret$\rightarrow$ret & pro$\rightarrow$ret& diff$\rightarrow$ret\\
\hline
No Feedback & 1830 & 1785 & 7 & 38 & 163 & 145 & 18 & 0\\
Weak Feedback & 1771 & 1596 & 47 & 128 & 302 & 212 & 57 & 33\\
Strong Feedback & 1779 & 1624 & 47 & 108 & 275 & 181 & 69 & 25\\
\hline
\end{tabular}
\caption{\label{proretrack} The angular momentum evolution of a sample of clouds between two adjacent data outputs. The columns are from left to right: the name of the simulation run, the total number of prograde clouds in the sample at the latter data output, the number of these prograde clouds that were prograde in the earlier output, the number of these prograde clouds that were retrograde in the previous output, the number of these prograde clouds that did not exist (i.e. that were part of the diffuse ISM) in the previous output, the total number of retrograde clouds in the sample at the latter data output, the number of these retrograde clouds that remained retrograde in the earlier output, the number of these retrograde clouds that were prograde in the previous output, and the number of these retrograde clouds that did not exist in the previous output.} 
\end{table*}

\subsubsection{General results}

Table~\ref{visctimes} shows the prograde rotation fractions at $t=500$ Myr for our three simulations. Our prograde fractions of $82-88$~\% are significantly larger than in the models of \citet{2011MNRAS.417.1318D} and \citet{2009ApJ...700..358T}, but similar to those of the models of \citet{2011ApJ...730...11T}. This is likely a result of our weak feedback, as well as our resolution limitations and our softening-length causing clouds to be more greatly influenced by large-scale galactic shear forces: our softening length of $60$ pc is larger than the minimum cell size of $7.8$ pc in the AMR simulations of \citet{2009ApJ...700..358T} and our gas mass resolution of $13000 M_\odot$ is coarser than the $2500 M_\odot$ in the SPH simulations of \citet{2011MNRAS.417.1318D}. Preliminary results from higher resolution simulations show a drop in the prograde fraction at $t=280$ Myr from $85\%$ in the moderate-resolution ($4\times10^5$ gas particles) ``strong feedback'' run presented here, to $70\%$ in a run with $10^6$ gas particles and the same $t_*$. Our initial population of clouds is also produced from smooth disc initial conditions that have not yet been perturbed by feedback, and this may also contribute to the mass of the oldest clouds.

As only a small number of clouds are strictly retrograde, we also show the fraction of ``strongly prograde'' clouds with angular-momentum axes within $30^\circ$ of the galaxy's angular momentum axis. Here we can see a clear dependence on feedback, with stronger feedback having a smaller strongly prograde fraction. This agrees with the results of \citet{2011MNRAS.417.1318D} who found that greater star formation leads to a smaller prograde fraction, but disagrees with the results of \citet{2011ApJ...730...11T}, who found that the addition of diffuse photoelectric heating increased the prograde fraction from $60$\% to $88$\%.

\begin{figure}
\includegraphics[width=1.\columnwidth]{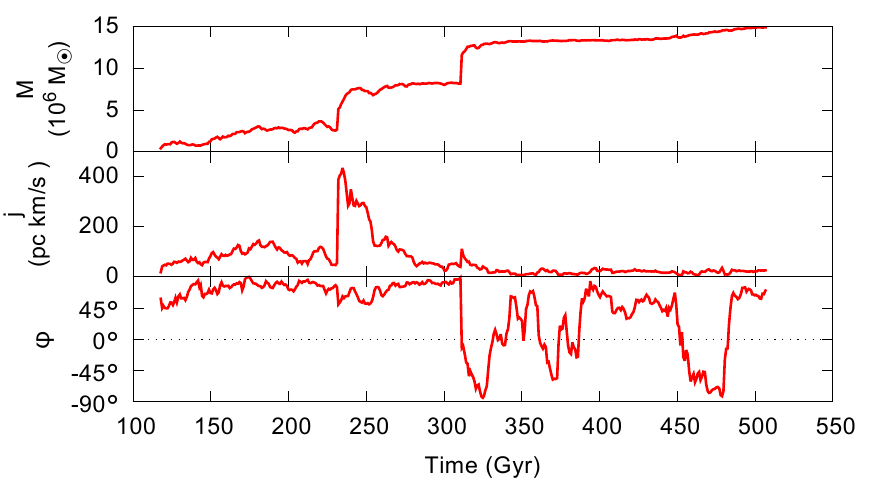}\\
\includegraphics[width=1.\columnwidth]{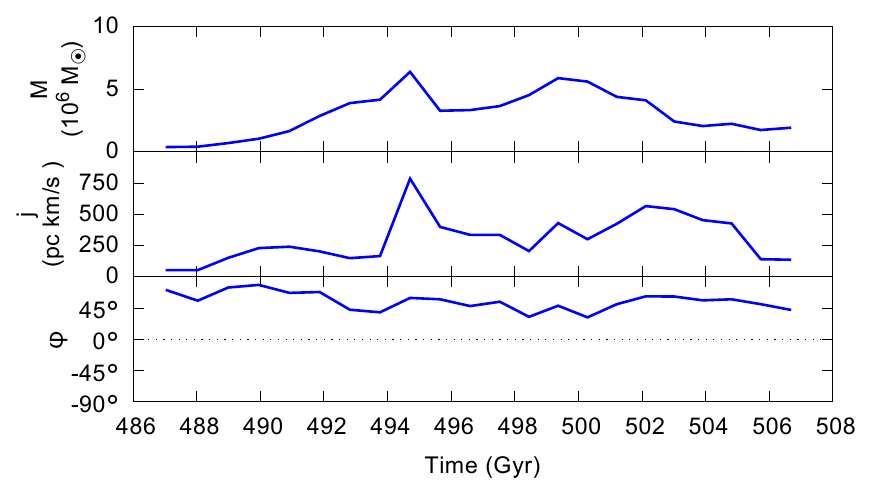}\\
\includegraphics[width=1.\columnwidth]{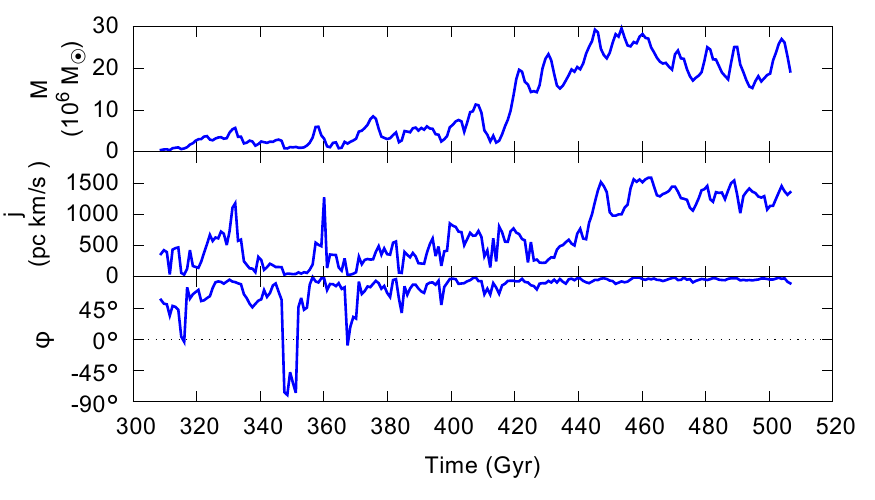}\\
\caption{\label{cloudsample} Tracks of mass $M$, specific angular momentum $j=L/M$, and spin axis $\phi$ (where $\phi=90^\circ$ represents purely prograde spin) of a selection of clouds. Top: Cloud A, a typically massive and long-lived cloud in the no-feedback run. Centre: Cloud B, a typically small and short-lived cloud in the strong-feedback run. Bottom: Cloud C, an unusually massive and long-lived cloud in the strong-feedback cloud.}
\end{figure}

\subsubsection{Typical cloud lives}

To illustrate the typical lives of clouds in our simulations, we identified individual clouds at $t=500$ Myr and tracked their primary progenitors, defined as the cloud in the previous output dump that contained the greatest number of gas particles from this cloud. The mass, specific angular momentum, and spin axis is plotted for the entire history of three selected clouds in Fig.~\ref{cloudsample}. In the no-feedback run, clouds are typically massive and long-lived. Cloud A (top) is an example of such a cloud. This cloud forms early in the simulation ($t\approx100$ Myr) as a low-mass strongly prograde cloud. The cloud's mass poceeds to gradually increase as mass is accreted. The cloud then undergoes a major merger at $t\approx230$ Myr, boosting its specific angular momentum. Another major merger occurs at $t\approx310$ Myr. This merger flips the spin of the cloud to retrograde motion. After this major merger the specific angular momentum is now low, and hence it is easy for events such as the loss and reintegration of material at the edge of the cloud, scattering events, and minor mergers, to dramatically change the cloud's spin axis, resulting in a spin axis that frequently switches between prograde and retrograde.

In the feedback runs, clouds are typically small and short-lived. Cloud B (Fig.~\ref{cloudsample}, centre) is a typical example of a small short-lived cloud in the strong-feedback run. Here the cloud is formed prograde from an evolved disc ($t=487$ Myr), gradually builds up mass and angular momentum, before dissipating itself after $20$ Myr. This cloud does not undergo any major interactions. Clouds similar to cloud B represent the majority of clouds in the feedback runs, and these clouds match the observed short life-times of molecular clouds.

Cloud C (Fig.~\ref{cloudsample}, bottom) is an unusually massive and long-lived cloud in the strong-feedback run. This cloud forms progade and small, but gradually begins to accumulate mass. At $t\approx350$ Myr the cloud loses a portion of mass, flipping its spin axis, until the gas is reintegrated. The cloud continues to merge and accumulate mass, but as the cloud is less bound, the energy input from a merger (both the kinetic energy of the collision and the thermal energy of the resulting star formation) disperses some fraction of the newly acquired gas. Each collision in cloud C therefore appears as a peak in the cloud mass, instead of a step as in cloud A. The cloud continues to collect mass and angular momentum, resulting in a strongly prograde cloud.

\subsubsection{Origin of cloud spins}
\begin{figure*}
\includegraphics[width=1.\columnwidth]{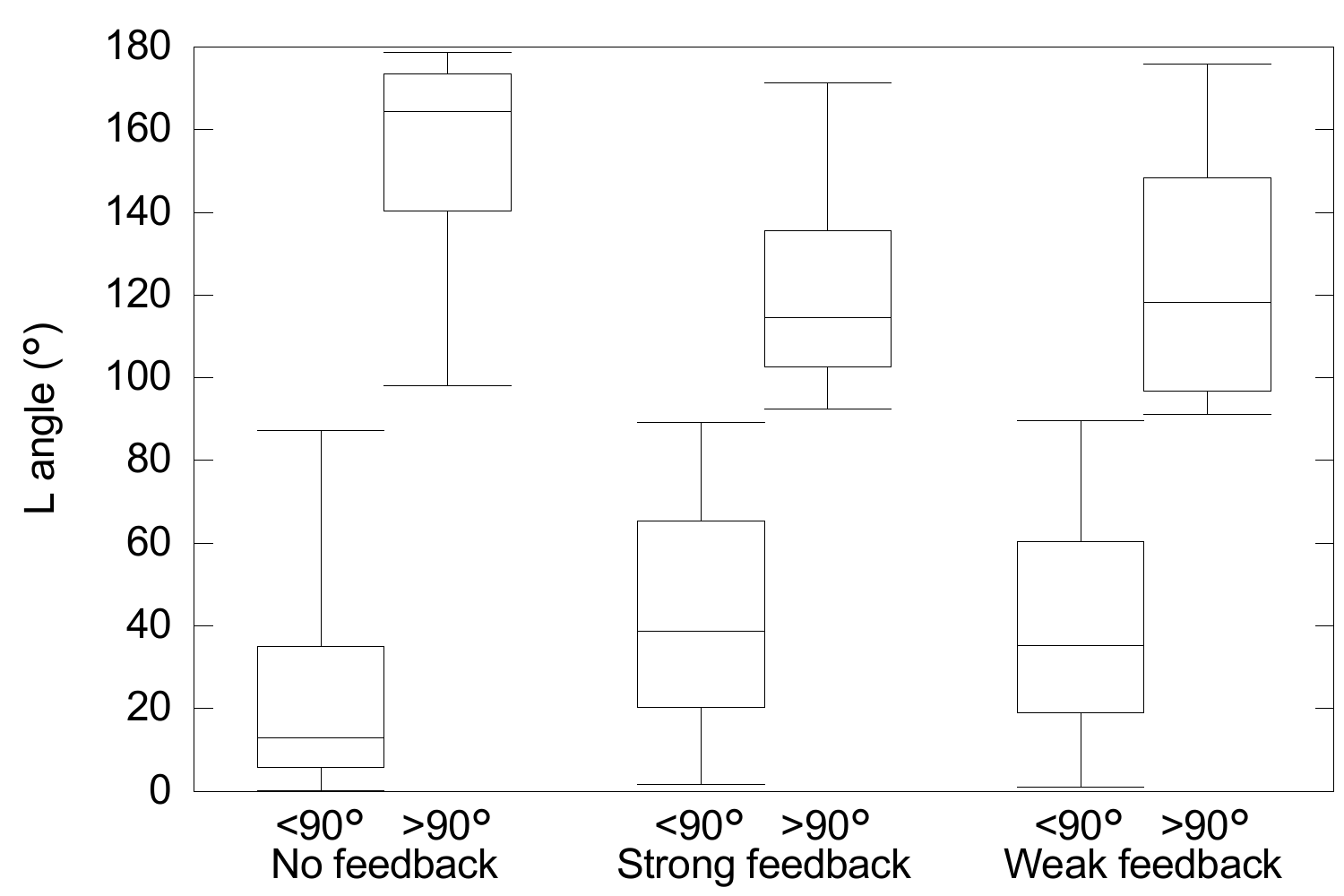}~
\includegraphics[width=1.\columnwidth]{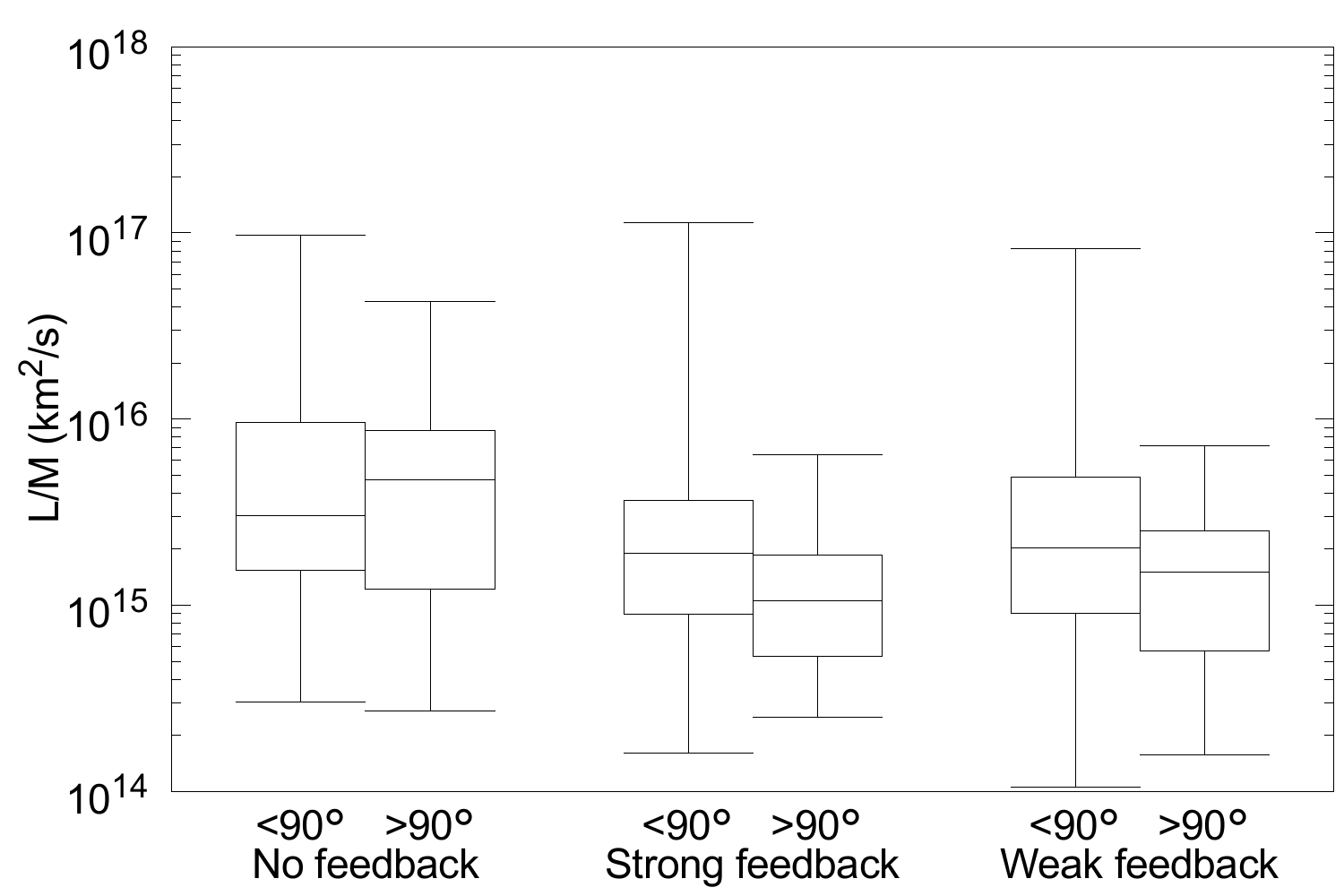}\\
\includegraphics[width=1.\columnwidth]{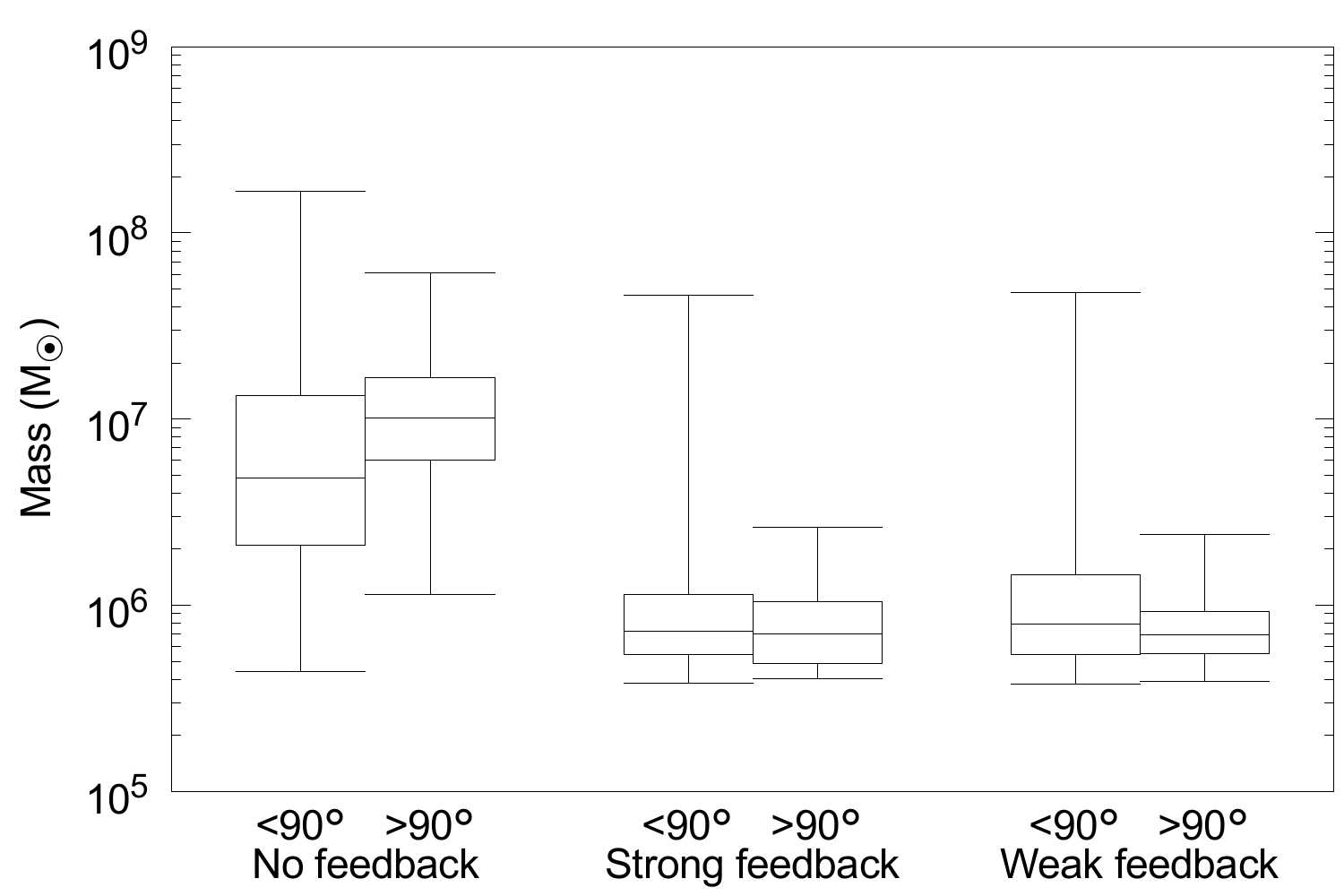}~
\includegraphics[width=1.\columnwidth]{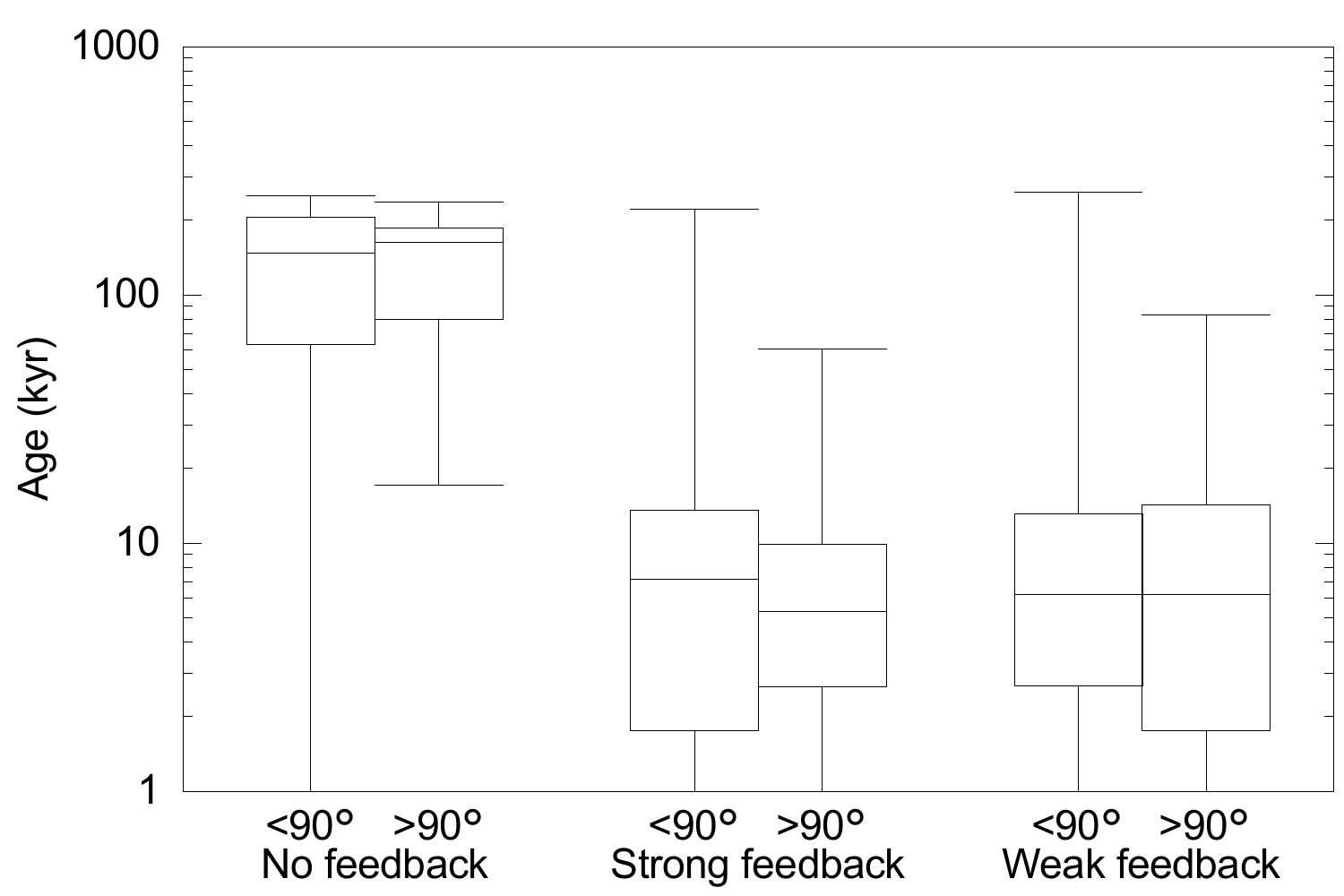}\\
\caption{\label{angrybins} Box and whisker plots, indicating the maximum, minimum, median, and upper and lower quartiles of the distributions of various cloud quantities as a function of feedback strength and cloud angular momentum, at $t=300$ Myr. The total size of each population is given in the caption of Fig.~\ref{nmerge}. Top left: $\theta$, the angle between the cloud's angular momentum axis and that of the galaxy. Top right: $L/M$ the magnitude of the cloud's specific angular momentum. Bottom left: $M$, the mass of the cloud. Bottom right: The age of the cloud in kyr.}
\end{figure*}

To explain the origin of angular momentum distribution of clouds, we binned the clouds according to angular momentum (prograde or retrograde), and identified these same clouds in the previous data dump (20 time-steps, or $\approx 1$ Myr earlier) to count how many clouds had changed angular momentum bins, how many clouds did not exist at the previous dump (i.e. they are new clouds formed from the diffuse ISM), and how many clouds stayed within the same bin. We applied this analysis to data dumps every $\approx 40$ Myr apart (1000 time-steps) for the first $440$ Myr. As the number of clouds in some of these categories is low at each time-step, we summed these results to improve the robustness of the statistics. These results are summarised in Table~\ref{proretrack}.

We found that in the feedback runs, over $100$ of the newly formed clouds were prograde (diff$\rightarrow$pro), while around $30$ of the newly formed clouds were retrograde (diff$\rightarrow$ret) - i.e. with feedback, about $20\%$ of newly formed clouds are retrograde. By contrast, without feedback, all new clouds were prograde (i.e. there are zero clouds in the diff$\rightarrow$ret category). This suggests that without feedback, retrograde motions must arise from encounters between clouds, but that with feedback, the clouds can form retrograde directly from the diffuse ISM.
With feedback, slightly more clouds changed from retrograde to prograde than vice versa, even though there are a lot more prograde clouds. Furthermore, a total of only $13-15$\% of the clouds in the feedback runs were retrograde (regardless of source, i.e. $N_\mathrm{retrograde}/(N_\mathrm{retrograde}+N_\mathrm{prograde})=0.13-0.15$), less than that of newly formed clouds. Overall, without feedback the retrograde population is only produced as clouds evolve and interact with each other and the ISM, but with feedback the clouds are formed both prograde and retrograde, and as clouds evolve and interact, a retrograde cloud has a greater chance of becoming prograde between dumps ($22-26$\%) than vice versa ($3.6-4.3$\%).

We also performed an analysis on clouds at a single time, $t=300$ Myr. We have placed these clouds in two bins according to angular momentum (prograde or not) to determine what differences there are in the populations. These data are plotted in Fig.~\ref{angrybins}. While there is considerable spread in the data, some trends are clear. In the no-feedback model the retrograde clouds are strongly retrograde and the prograde clouds are strongly prograde, while in the simulations with feedback the distribution of angular momenta is broader. This again suggests that in the absence of feedback, clouds change their angular momentum more dramatically, i.e. with strong scattering events. In all simulations, the most massive clouds are prograde, which is not surprising as the most massive clouds will sample a significant portion of the galaxy's rotation curve, and should be strongly prograde. With the inclusion of feedback, the majority of retrograde and prograde clouds have a similar mass. This contrasts to the no feedback run where the retrograde clouds are more massive. This is likely because retrograde clouds in the no-feedback run are only formed by collisions and mergers. Clouds in the no-feedback run are much older than in the feedback runs, as they are not disrupted. Without feedback, retrograde clouds are at least $20$ Myr old, implying they must live long enough to experience a collision that flips the angular momentum axis. With feedback, prograde and retrograde clouds have similar ages, although there exists a small number extremely old, massive, and strongly prograde clouds.

\begin{figure}
\includegraphics[width=1.\columnwidth]{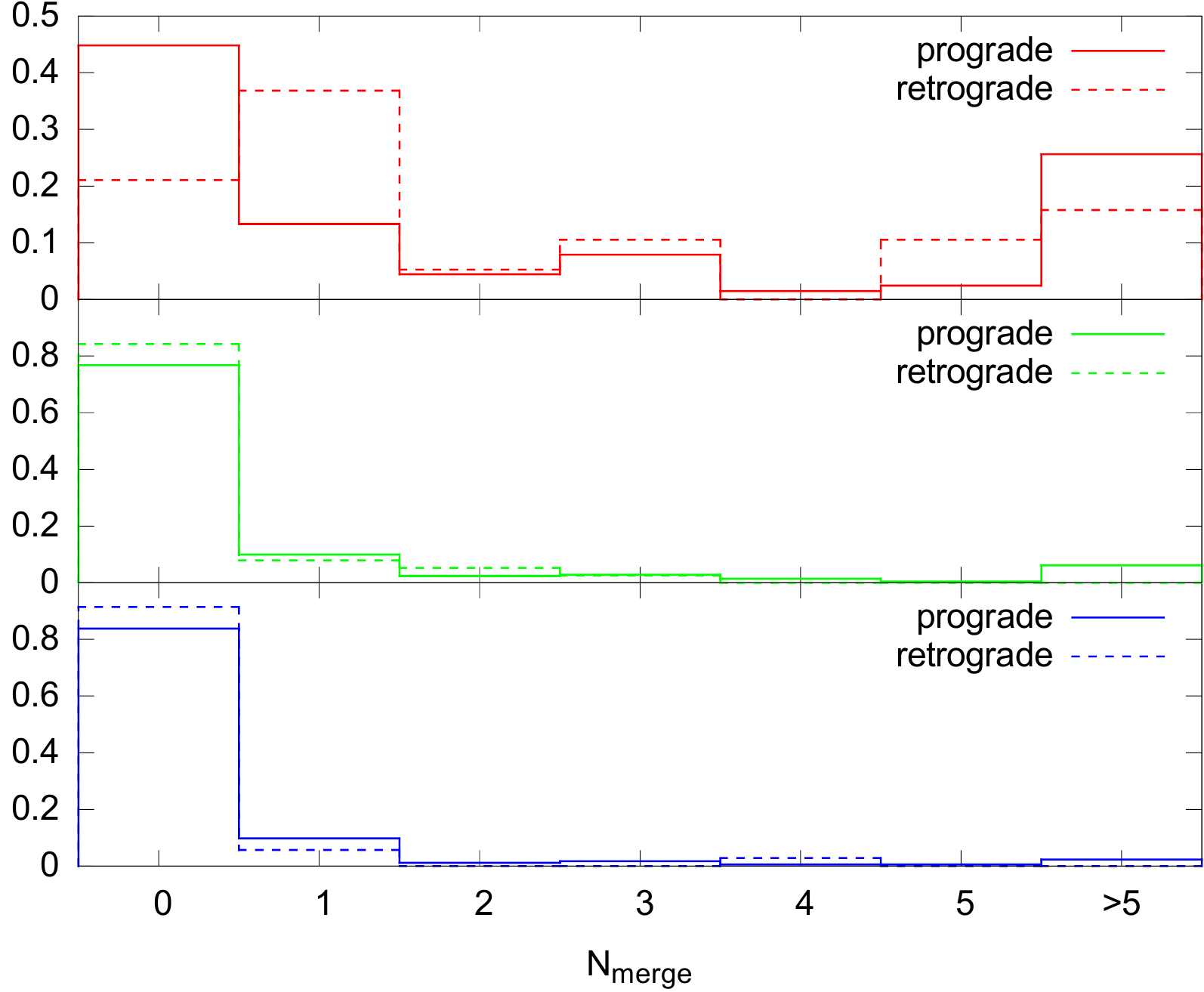}
\caption{\label{nmerge} Histograms for the number of mergers experienced by the prograde and retrograde populations at $t=300$ Myr. Top: no feedback. Centre: weak feedback. Bottom: strong feedback.  At this time, the no-feedback run has $203$ prograde clouds and $19$ retrograde clouds, the strong-feedback run has $173$ prograde clouds and $35$ retrograde clouds, and the weak-feedback run has $211$ prograde clouds and $38$ retrograde clouds.}
\end{figure}

We also counted the number of mergers each cloud at $t=300$ Myr had previously experienced (Fig.~\ref{nmerge}). In the no feedback run, a significant fraction of both prograde and retrograde clouds underwent a large number of mergers. Without feedback, clouds are long-lived and hence clouds may undergo many collisions. Almost half of the prograde clouds have undergone no mergers at all, while approximately $80$\% of retrograde clouds have undergone at least one collision, further suggesting that retrograde spin traces its origin to collisions between clouds in the no-feedback model. With the inclusion of feedback, the retrograde and prograde populations are similar to each other. However, a slightly larger fraction of prograde clouds have undergone a collision, suggesting that collisions between clouds may be slightly biased in favour of producing more prograde clouds. This may be because as clouds merge, their combined angular momentum contains a greater fraction of the disc's mass, and thus is slightly more likely to be representative of the large-scale rotation curve of the disc, and slightly less affected by random small-scale velocity deviations.

Our proposed picture is as follows: when clouds form, they are small in size and hence their angular momentum is drawn from the small scale velocity field of the ISM. In the presence of feedback, the velocity field of the ISM has a significant component of random motions on these scales, and so the angular momentum vector of an individual cloud is somewhat random, even though the average angular momentum of a large number of clouds must necessarily be prograde. As a cloud merges with more clouds or accretes additional matter, this randomness is averaged away and the larger scale shear starts to dominate, resulting in more prograde clouds. However, in simulations without feedback, clouds are initially prograde, and can only become retrograde through collisions. Hence while violent collisions can also flip clouds, they are likely not the main cause of retrograde rotation in a realistic turbulent ISM: indeed mergers between clouds act to reduce retrograde rotation.

Scattering events and violent mergers of clouds are not likely the main cause of retrograde rotation in simulations with feedback, as then we would expect clouds to become increasingly retrograde over time -- i.e. we would expect prograde clouds to be younger or less massive as in the no-feedback runs. Scattering and violent mergers could indirectly contribute to retrograde rotation by stirring up the diffuse ISM from which clouds form, but localised stellar feedback is the primary source of turbulence. However, smooth global feedback \citep[as in][]{2011ApJ...730...11T} will reduce the Reynolds number of the diffuse gas by heating it, and hence the clouds that form from this less-turbulent ISM are more strongly prograde.

\section{Conclusions}\label{concsect}

We have performed simulations to determine the effects of stellar feedback on the formation and evolution of giant molecular clouds. We produced algorithms to identify clouds and analyse their properties. We make the following conclusions:
\begin{itemize}
\item We find that the viscous time-scale due to cloud-cloud collisions decreases with the addition of feedback. The clouds are less massive and collisions between them are more frequent when feedback is included, because cloud collisions are less violent and less efficient at losing energy. 

\item We also find that the feedback algorithm considered here significantly reduces the number of clouds with strongly prograde rotation. After careful analysis, we conclude that small young clouds are more strongly influenced by the turbulence of the ISM and are more likely to form retrograde while large old clouds tend to approach the average angular momentum of the galaxy and thus are more likely to be prograde. Stellar feedback contributes to turbulence and disrupts clouds (reducing the population of old large clouds), producing fewer strongly prograde clouds.

\item Finally, we find that interactions between clouds produce very different results depending on the presence of feedback. Without feedback, interactions primarily act to increase the retrograde fraction through scattering events between clouds. However when localised feedback is included, interactions have little effect, and perhaps act to increase the {\em prograde} fraction as clouds form from a high velocity-dispersion medium, and only later merge and grow in mass and angular momentum. Diffuse heating does not have the same effect as localised feedback, as it acts to smooth out the density distribution and rotation curve, producing more strongly prograde clouds.
\end{itemize}

\section*{Acknowledgements}

We thank Larry Widrow for providing his initial conditions generator. The Friends-of-Friends code we used was downloaded from the University of Washington N-body shop {\tt 
http://www-hpcc.astro.washington.edu/}. This research is supported by the Canada Research Chairs Program and NSERC. RJT is also supported by the Canada Foundation for Innovation, and the Nova Scotia Research and Innovation Trust. JW was supported by Saint Mary's University while the simulations were performed, and is currently supported by Monash University. BKG acknowledges the support of the UK's Science $\&$ Technology Facilities Council. Simulations were run on 
the CFI-NSRIT funded {\em St Mary's Computational Astrophysics Laboratory}.
\bibliographystyle{mnras}
\bibliography{cloud_feedback}

\bsp

\label{lastpage}

\end{document}